\documentclass[aps,prl,twocolumn,preprintnumbers,nofootinbib,superscriptaddress,longbibliography]{revtex4-1}

\usepackage{amsmath,amssymb,mathtools,bm}
\usepackage{graphicx, color, hepunits}
\usepackage[dvipsnames]{xcolor}
\usepackage{float}
\usepackage[normalem]{ulem}
\usepackage{filecontents}
\usepackage{hyperref} 
\hypersetup{
    colorlinks=true,       
    linkcolor=blue,        
    citecolor=blue,        
    filecolor=magenta,     
    urlcolor=blue          
}
\usepackage[utf8]{inputenc}
\usepackage[english]{babel}

\usepackage{cleveref}
\usepackage[toc,page]{appendix}

\AtBeginDocument{%
    \newwrite\bibnotes
    \def\bibnotesext{Notes.bib}
    \immediate\openout\bibnotes=\jobname\bibnotesext
    \immediate\write\bibnotes{@CONTROL{REVTEX41Control}}
    \immediate\write\bibnotes{@CONTROL{%
    apsrev41Control,author="08",editor="1",pages="1",title="0",year="1"}}
     \if@filesw
     \immediate\write\@auxout{\string\citation{apsrev41Control}}%
    \fi
}%

\newcommand{\order}[1]{\mathcal{O}(#1)}

\begin{document}
\title{Superconducting Levitated Detector of Gravitational Waves}
\author{Daniel Carney}
\affiliation{Physics Division, Lawrence Berkeley National Laboratory, Berkeley, CA}
\author{Gerard Higgins}
\affiliation{Institute for Quantum Optics and Quantum Information (IQOQI),
Austrian Academy of Sciences, Vienna, Austria}
\affiliation{Institute of High Energy Physics (HEPHY),
Austrian Academy of Sciences, Vienna, Austria}
\author{Giacomo Marocco}\thanks{gmarocco@lbl.gov}
\affiliation{Physics Division, Lawrence Berkeley National Laboratory, Berkeley, CA}
\author{Michael Wentzel}\thanks{wentzel4@illinois.edu}
\affiliation{Department of Physics, University of Illinois Urbana-Champaign, Urbana, IL}
\preprint{ }

\begin{abstract}
A magnetically levitated mass couples to gravity and can act as an effective gravitational wave detector. We show that a superconducting sphere levitated in a quadrupolar magnetic field, when excited by a gravitational wave, will produce magnetic field fluctuations that can be read out using a flux tunable microwave resonator. With a readout operating at the standard quantum limit, such a system could achieve broadband strain noise sensitivity of $h \lesssim 10^{-20}/\sqrt{\rm Hz}$ for frequencies of $1~\mathrm{kHz}~-~1~\mathrm{MHz}$, opening new corridors for astrophysical probes of new physics.
\end{abstract}

\maketitle


Gravitational waves have been detected in the Hz--kHz regime with laser interferometers \cite{LIGOScientific:2016aoc}, and potentially in the nHz regime with pulsar timing array observations \cite{NANOGrav:2023gor}. Beyond these important frequency ranges, an array of potential signals exists across the frequency spectrum, including signatures from cosmology, astrophysics, and a variety of speculative physics Beyond the Standard Model. Correspondingly, a number of methods for detecting gravitational waves in the low-frequency nHz--Hz~\cite{Janssen:2014dka,Namikawa:2019tax,CMB-S4:2020lpa,Badurina:2019hst,Abe:2021ksx,AEDGE:2019nxb,Hild:2010id, Punturo:2010zz,LIGOScientific:2016wof,LISA:2017pwj,Yagi:2011wg,Gulian,ISHIDOSHIRO20101841} and high-frequency $\gtrsim$~kHz~\cite{Holometer:2016qoh,Goryachev:2021zzn,Aggarwal:2020olq,Aggarwal:2020umq,Berlin:2021txa,Domcke:2022rgu,Berlin:2023grv, Bringmann:2023gba, Domcke:2023bat} regimes are in various stages of development. 

In this paper, we suggest a superconducting levitated detector of gravitational waves (SLedDoG) as a broadband method to detect gravitational waves (GWs) in the 1 kHz--1 MHz regime. This regime is particularly motivated by a number of potential signals, including 
astrophysical signatures of physics Beyond the Standard Model~\cite{Brito:2015oca,Arvanitaki:2010sy,Arvanitaki:2014wva}, GWs sourced from mergers of neutron stars~\cite{Casalderrey-Solana:2022rrn}, and light primordial black holes~\cite{Franciolini:2022htd,Franciolini:2022ewd,hooper2020hot}. 

\begin{figure}[t!]
    \centering
    \includegraphics[width=0.48\textwidth]{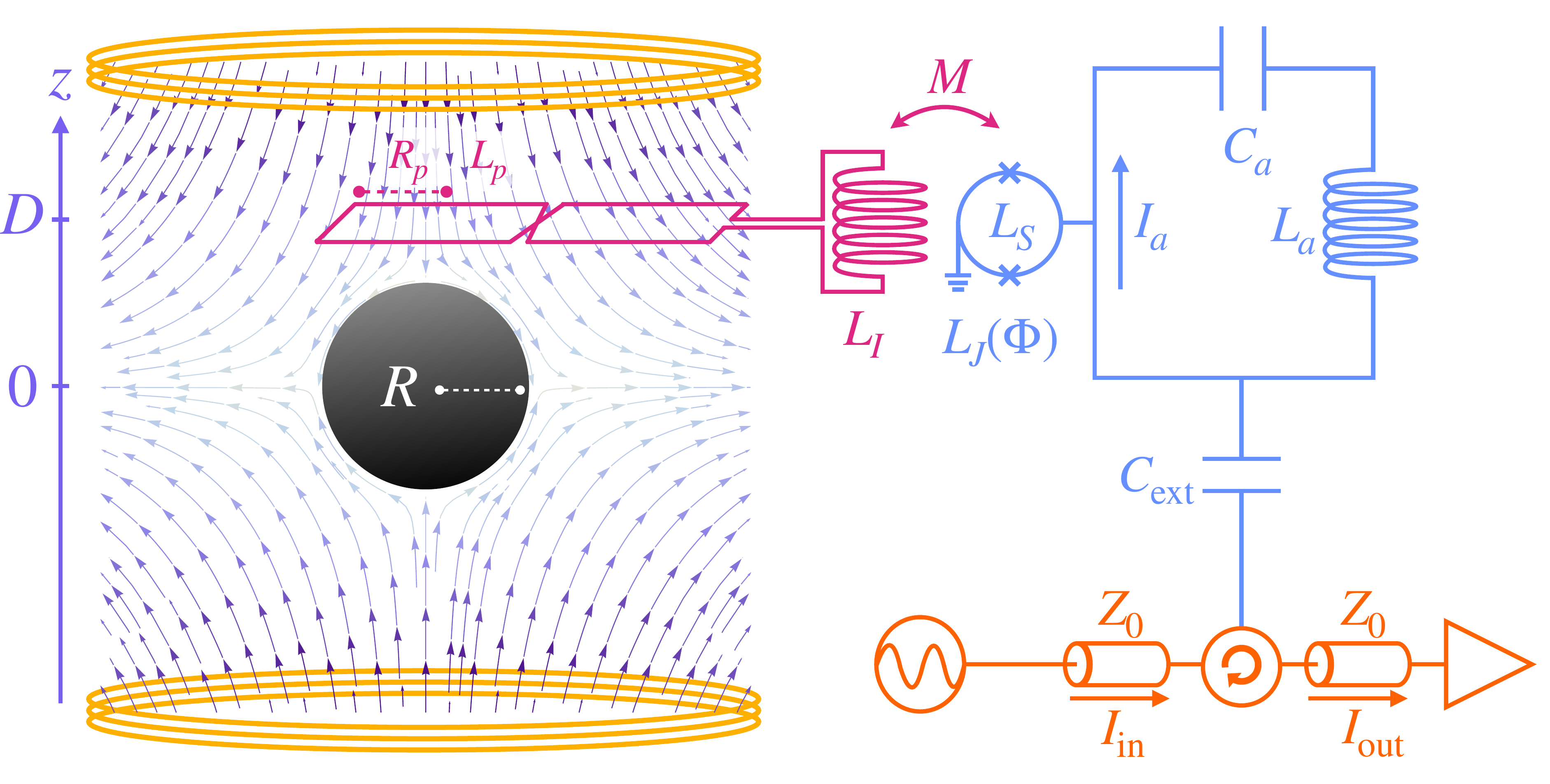} \\
    \includegraphics[width=.48\textwidth]{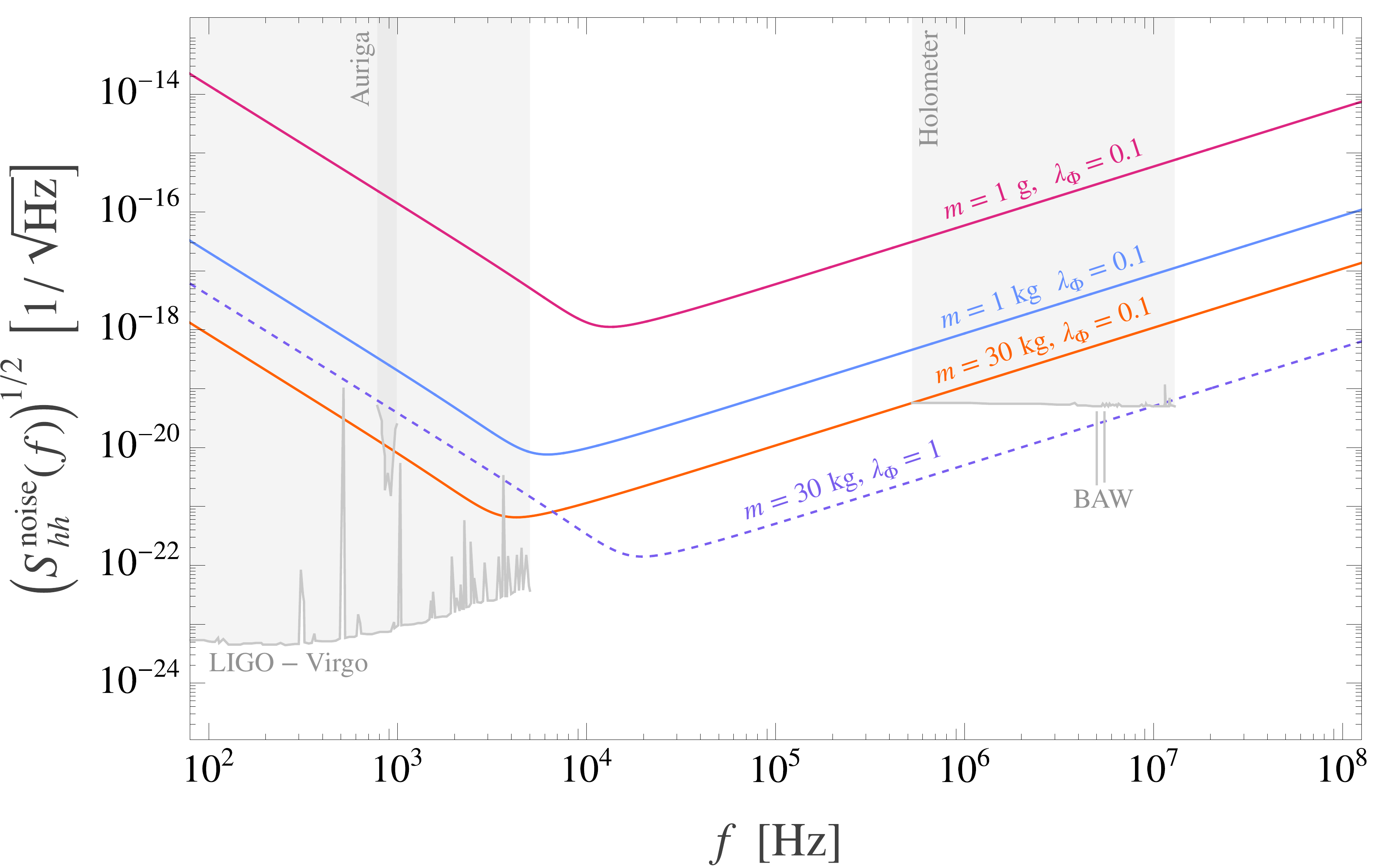}
    \caption{Top: basic SLedDoG detector concept with a superconducting sphere levitated by anti-Helmholtz coils (gold), gradiometric pickup and input coils (magenta), microwave resonator (blue), and microwave drive (orange). Bottom: predicted strain  sensitivity. Solid curves show the sensitivity for $1$~g, $1$-kg, and $30$-kg setups with $\lambda_{\Phi} = 0.1$, corresponding to sphere radii of $2.7$~mm, $2.7$~cm, and $8.6$~cm respectively. The dashed curve shows the sensitivity for the $30$~kg setup with $\lambda_{\Phi} = 1$. The sensitivity is optimized at the maximum achievable SQL frequency $\omega_*$ for each setup (see Appendix B). We take $\kappa = 2 \omega_*$, $\beta = 1.6$, $T = 10$~mK, $\rho = \rho_{\rm Pb}$, $B_c = 1$~T, $D = R$, $\bar{n} = 100$, $\omega_a = 10~\text{GHz}$, and $d\omega_a/d\Phi = 2\pi\times 1~\mathrm{GHz}$.  The shaded regions depict the strain sensitivities of existing experiments.}
    \label{fig:setup}
\end{figure}

The detector concept and predicted reach are shown in Fig. \ref{fig:setup}. A quadrupolar magnetic field is used to levitate a superconducting test mass. An incoming gravitational wave causes a pickup coil to move relative to the sphere, leading to a time-dependent magnetic flux through the coil. This flux drives a current; continuous measurement of this current leads to continuous measurement of the sphere position. Similar platforms are being developed for tests of quantum physics with sizable masses \cite{romero-isartQuantum2012, Waarde2016, GutierrezLatorre2022, GutierrezLatorre2023, PhysRevLett.131.043603, Schmidt2024}, dark matter searches \cite{higginsMaglev2023} and gravity gradiometers \cite{PhysRevApplied.8.064024}. The current can be probed using superconducting quantum interference devices (SQUIDs) or flux tunable microwave resonators (FTMR) \cite{Schmidt2024,Schmidt2020,Rodrigues2019,Bothner2021,richman2023general}, such as a radio frequency quantum upconverter \cite{kuenstner2022quantum}. Here, we focus on the FTMR approach, since, as we will show, it allows for broadband sensing at high frequencies.


The essence of our proposal is similar to GW detection with a laser interferometer, where an optical field is used to continuously monitor the distance of a nearly freely-falling test mass with respect to a reference. In modern incarnations of these detectors, the dominant noise source in the detection band comes from the quantum noise in the optical readout, roughly at the level of the Standard Quantum Limit (SQL)~\cite{caves1981quantum,beckey2023quantum}. Because the coupling of a single photon to the test mass motion is weak, reaching the SQL at higher frequencies in an optical system quickly becomes prohibited by the need for stronger lasers. In contrast, as we emphasize below, magnetic systems can achieve orders-of-magnitude larger couplings to the individual microwave photons in a superconducting readout circuit. This could enable operation of a detector at the SQL at frequencies well above the audio band.

\textit{Detector concept}\label{sec:setup}--- We begin by discussing the essential physics of the levitating sphere and the interaction of the system with gravitational waves, before characterizing the readout system. Consider a superconducting sphere of radius $R$ and density $\rho$ placed in a quadrupolar magnetic field. In the presence of the magnetic field, current loops form on the surface of the sphere to expel magnetic field lines inside the bulk. This system exhibits a stable equilibrium at the center of the quadrupolar magnetic field. For a quadrupolar magnetic field $\mathbf{B}_0 = \frac{b_z}{2}\left(x\hat{\mathbf{x}} + y\hat{\mathbf{y}} - 2z\hat{\mathbf{z}}\right)$ centered on the origin with $\partial B_z/\partial z = -b_z$, the sphere is harmonically trapped with angular frequency
\begin{equation}
\label{eq:omega0}
\omega_0^2 = \frac{3 b_z^2}{2 \rho}
\end{equation}
and is insensitive to the sphere size provided the radius is much larger than the penetration depth of the superconductor~\cite{Hofer_2019}. For reference, a lead sphere in a quadrupole field with gradient $b_z = 29~\mathrm{T}/\mathrm{m}$, corresponding to the blue curve in Fig.~\ref{fig:setup}, 
has resonant frequency $f_0 = \omega_0/2\pi \approx 47~\mathrm{Hz}$.

The trapping magnetic quadrupole field can be produced using current-carrying coils in an anti-Helmholtz configuration. A gradiometric pick-up loop oriented in the $xy$-plane measures the flux induced by motion of the levitated sphere, while being insensitive to motion of the trapping coils due to the approximate invariance of the $z$-component of the trapping field under $xy$-translations, as discussed in Appendix A.

Now consider a GW incident on this system, with wavelength much longer than the size of the system. We will describe the signal in the so-called ``proper detector frame'' \cite{manasse1963fermi, Maggiore}, in which the metric is locally flat at the origin, taken to be the center of the quadrupole field. In this frame, the elements of the apparatus feel a force due to the GW 
\begin{align}
    F^\mathrm{GW}_i(\mathbf{x}) = \frac{m}{2} x^j \ddot{h}^{TT}_{ij}
    \label{eq:F}
\end{align}
that depends on the position $\mathbf{x} = (x, y, z)$ and mass $m$, as well as the metric $h$, evaluated in transverse-traceless gauge~\cite{Maggiore_2007}. We take $m$ to be the mass of the sphere, assuming that the pickup is attached to a heavy apparatus. The GW force \eqref{eq:F} induces a change $\xi$ in the distance $D$ between the sphere and the pickup loop; with our parameters, the distance between the trap center and sphere is negligible. In the limit that the GW frequency is large enough that the motion of $\xi$ is approximately free, $\xi$ obeys
\begin{align}
 \ddot{\xi} \approx  \frac{\ddot{h}_{zz}}{2} D   ~.
 \label{eqn:signal}
\end{align}
This change in displacement causes a change in flux through the pick-up loop, which we read out as discussed below.

In addition to causing relative motion between the sphere and the pick-up loop, the GW also moves the primary coils and distorts the shape of the elements of the experimental apparatus. In the frame with the origin at the trap center, the two primary coils feel equal and opposite forces in the $z$-direction. For simplicity, in the rest of the paper we focus on +-polarised GWs incident along the $xy$-plane and relegate a complete discussion of the other GW-induced strains to Appendix A. For such a GW, the GW signal comes from changes in $\xi$, and the phenomena from other GW configurations affect the signal by only an $\mathcal{O}(1)$ amount compared to this expectation.

Proposals to use levitated superconductors as dark matter detectors and gravity gradiometers use SQUIDs to read out changes in the magnetic field~\cite{higgins2023maglev,thewindchimecollaboration2022snowmass,antypas2022new,PhysRevApplied.8.064024,richman2023general}. However, at frequencies higher than $\sim 10$~kHz, it becomes difficult to strongly couple the SQUID to the displacement of the sphere, leading to large imprecision noise. We instead show that a flux tunable microwave resonator (FTMR, \cite{kuenstner2022quantum,Schmidt2024}) can strongly couple to the displacement of the sphere at high frequencies. The 
FTMR, pictured in Fig. \ref{fig:setup}, is a driven microwave resonator, terminated via a SQUID, inductively coupled to displacement of the sphere via a pickup circuit --- a transformer which transfers flux from a pickup coil to the SQUID via an input coil, in which both the pickup and the input coil have inductances $L_p$. The microwave resonator is an LC circuit with capacitance $C \equiv C_a + C_{\rm ext}$ and inductance $L(\Phi) \equiv L_a + L_S + L_J(\Phi)$ dependent on the flux threading the SQUID, which depends on the separation of the pickup loop and the sphere. In a typical Josephson system, $L_J(\Phi)$ is periodic; we will be interested in the regime of small fluxes, where we can linearize $L_J(\Phi) \approx L_J(0) + (dL_J/d\Phi) \Phi$. In this regime, the resonator's frequency is also a function of the threading flux, which can be approximated as  $\omega_a(\Phi) \approx \omega_a + (d\omega_a/d\Phi) \Phi$. 

The microwave resonator dynamics and the mechanical motion of the sphere both amount to simple harmonic oscillators in these approximations. Let $a^\dag$ be the creation operator for a microwave resonator photon. The Hamiltonian $H_{\rm LC} = \omega_a(\Phi) a^\dag a$ then has frequency depending on the magnetic flux threading the resonator, as discussed above. For small mechanical displacements $\xi$, this flux is
\begin{equation}
    \label{eq:flux-from-displacement}
    \Phi = \beta \lambda_{\Phi} R^2 b_z \xi \equiv \eta \xi~,
\end{equation}
where $\lambda_{\Phi} \approx M/2L_p$ characterizes the transduction between flux through the pickup loop $\Phi_p$ and the readout circuit $\Phi$ in terms of the mutual inductance $M$ of the input coil and the SQUID, $\beta$ is a dimensionless parameter dependent on the geometry of the pickup loop and the trapping field $\mathbf{B}_0$~\cite{hofer2022highq,hofer2024numerical}, and $R$ is the radius of the sphere. After linearizing around $\xi = 0$, the Hamiltonian $H_{\rm sys}$ for the sphere and resonator can be written as
\begin{equation}
    \label{eq:system-hamiltonian-linear}
    H_{\rm sys} = \omega_a a^\dag a + \omega_0 b^\dag b - G_0 a^{\dagger} a \left( b + b^{\dagger}\right)~.
\end{equation}
Here $\xi = \xi_0 \big(b + b^{\dagger}\big)$ where $b^{\dagger}$ is the creation operator for the sphere center of mass motion, and we have defined the zero point fluctuation scale and single-photon coupling
\begin{equation}
\label{eq:G0}
\xi_0 = \frac{1}{\sqrt{2 m \omega_0}}, \ \ \ G_0 = \eta \xi_0 \frac{d\omega_a}{d\Phi}.
\end{equation}

The resonator is driven by a source of microwave photons with frequency $\omega_d$~\cite{kuenstner2022quantum,Fedorov_2016}. The drive photons pick up a phase shift as they circulate the resonator and then exit, where their phase can be measured, again in analogy with the laser photons in an interferometer. The drive effectively enhances the single-photon coupling $G_0$. To model both of these effects, we employ the standard input-output formalism~\cite{beckey2023quantum}. The output charge is related to the input charge through the usual I/O relation
\begin{equation}
    \label{eq:boundary-condition-j}
    q_{\text{out}} = q_{\text{in}} - \sqrt{\kappa} q,
\end{equation}
where $\kappa$ is the loss rate of the microwave resonator, taken to be dominated by the measurement port loss. Linearizing around the drive and accounting for possible thermal noise, the Heisenberg-Langevin equations of motion for the system become
\begin{align}
\begin{split}
\label{eq:EOM_main}
\dot{\xi} & = \frac{p}{m} \\
\dot{p} & = - m \omega_0^2 \xi - \frac{G}{\phi_0^2 \xi_0} \phi -    \gamma p + F_{\rm in} \\
\dot{\phi} & = \frac{q}{C}- \kappa \phi + \sqrt{\kappa} \phi_{\rm in}~ \\
\dot{q} & = - C \omega_a^2 \phi  - \frac{2 G}{\phi_0^2 \xi_0} \xi - \kappa q + \sqrt{\kappa} q_{\rm in}~.
\end{split}
\end{align}
Here, $q$ and $\phi$ are the canonically conjugate charge and flux in the FTMR, $\gamma$ is the mechanical loss rate, $F_{\rm in}$ is the input operator for thermal noise, and $G = G_0 \sqrt{\overline{n}}$ is the single-photon coupling enhanced by the presence of $\overline{n} \gg 1$ circulating photons in the microwave resonator. In practice, $\overline{n}$ is limited by the requirement that the inductance $L_J(\Phi)$ be linear in $\Phi$. A system with a large $G$ is strongly coupled, such that small fluctuations in the position of the sphere will lead to large fluctuations of the output modes.

The equations of motion are easily solved in frequency domain. This gives the output current, our basic observable:
\begin{align}
\begin{split}
    \label{eq:signal-mode}
    q_{\rm out} & = e^{i \phi_c} q_{\rm in} + 2 \left( \frac{G}{\phi_0 \xi_0} \right)^2 \kappa \chi_c^2 \chi_m \phi_{\rm in} \\
    & + 2 \left( \frac{G}{\phi_0 \xi_0} \right) \sqrt{\kappa} \chi_c \chi_m m \nu^2 D h_{\rm in}.
\end{split}
\end{align}
Here, $h_{\rm in}$ is the gravitational wave signal [which gives a force $F_{\rm in} = m D \nu^2 h_{\rm in}/2$, see Eq. \eqref{eq:F}], and the LC circuit (``cavity'') and mechanical motion susceptibilities are
\begin{equation}
    \label{eq:cavity-response-function}
    \chi_c(\nu) = \frac{1}{i \nu -  \kappa/2}~, \ \ \chi_m(\nu) = \frac{1}{m [(\omega_0^2 - \nu^2) - i \gamma \nu]},
\end{equation}
and the phase $e^{i \phi_c} = \chi_c/\chi_c^*$. The second line of Eq. \eqref{eq:signal-mode} shows the gravitational wave is encoded into the measured output on the microwave line. The first line shows the two noise sources added by the readout itself: the shot noise, encoded by the $q_{\rm in}$ field, and the back-action noise, encoded by the $\phi_{\rm in}$ field. In the next section, we analyze the relative size of these terms and quantify their added noise.

\begin{figure}[t]
\centering
\includegraphics[width=.48\textwidth]{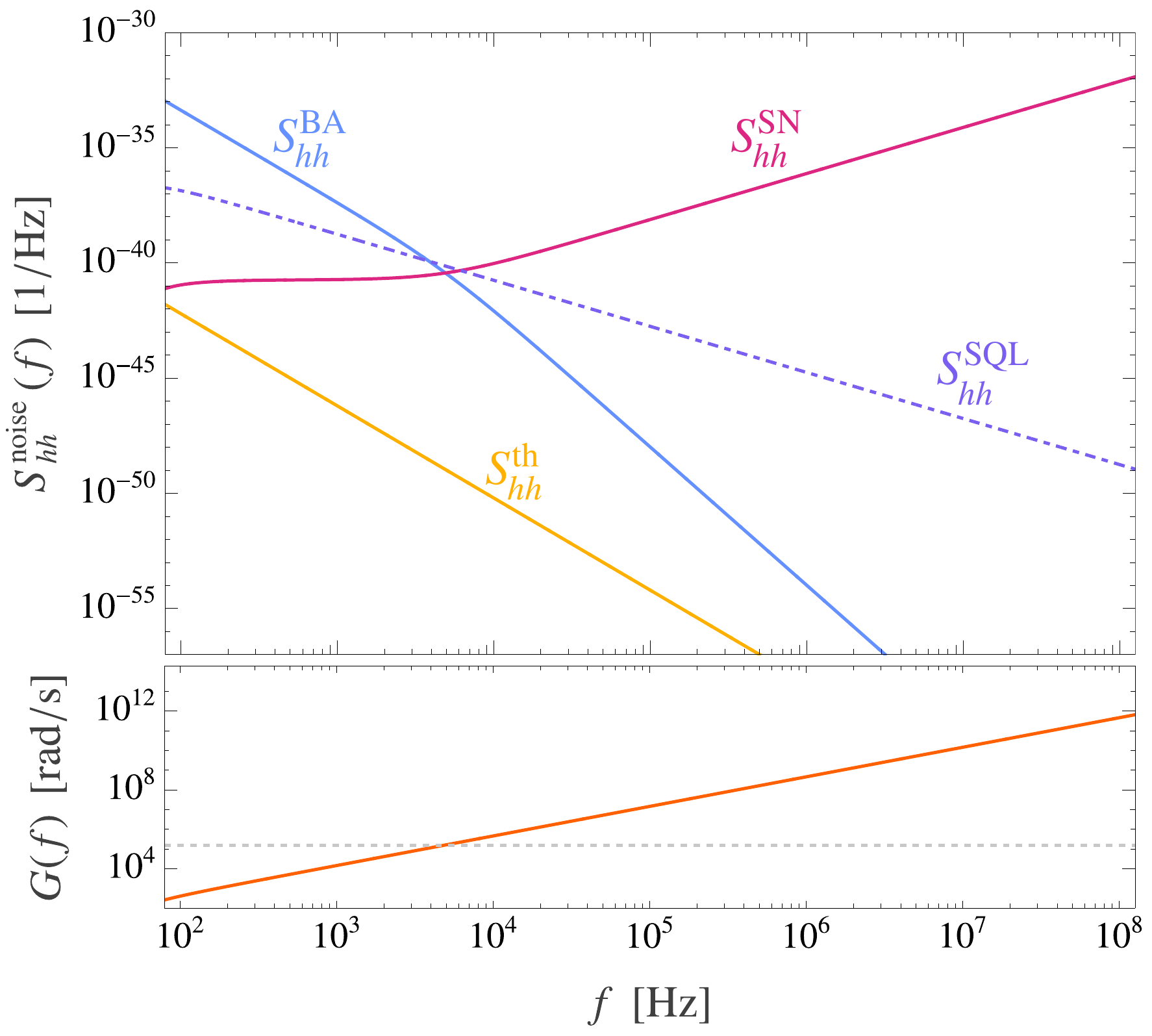}
\caption{Noise PSDs referred to the GW strain for a $1$~kg SLedDoG setup with $\lambda_{\Phi} = 0.1$. Solid lines show noise profiles for a setup reaching the SQL at $\omega_* = \kappa/2 \approx 2\pi\times 23$~kHz. The dashed line shows the SQL, where $S_{hh}^{\rm SN} = S_{hh}^{\rm BA}$. We note that at $T = 10$~mK, the vibrational modes of the sphere will be thermally occupied leading to narrow peaks in the thermal noise PSD near mechanical resonances for $f \gtrsim 10$~kHz. In practice, these frequencies will be excised from a physics analysis, which should only affect small bandwidths of the experiment. The bottom panel shows the coupling $G(\nu)$ required to reach the SQL at frequency $\nu$ for a $1$~kg setup (solid) and the estimated achievable coupling $G$ (dashed), whose value is discussed around Eq. \eqref{eq:G0-numbers}.}
\label{fig:noise-profiles}
\end{figure}

\textit{Noise and sensitivity}
\label{sec:noise}--- We assume that our detector is limited by irreducible noise from the surrounding thermal bath, as well as the quantum measurement noise added from the readout system. In practice, operating the microwave LC circuit with $\omega_a = 10~\text{GHz}$ in a $T \sim 10~{\rm mK}$ fridge we have $\omega_a \gg T$. This means that the readout circuit's uncertainty is dominated by its quantum vacuum noise. The thermal noise on the mechanics is not important in this regime for a reasonably high-$Q$ mechanical system, as we show quantitatively at the end of this section; see also Fig. \ref{fig:noise-profiles}. 

The strain noise power spectrum of the detector can be computed using the solution \eqref{eq:signal-mode} and the Weiner-Khinchin theorem, following standard techniques \cite{beckey2023quantum}. We also provide a detailed derivation in Appendix B. The result takes the form $S_{hh} = S^{\rm T}_{hh} + S^{\rm SN}_{hh} + S^{\rm BA}_{hh}$, where the first term represents thermal noise acting on the sphere, and the last two terms are the quantum readout noise, called shot noise and back-action respectively. 

The total quantum readout noise can be minimized at a single frequency of choice. The shot noise term is due to vacuum fluctuations in the phase of the microwave drive, while the back-action is from the random inductive force acting on the mechanical system due to the microwave drive fluctuations. Explicitly,
\begin{align}
\begin{split}
\label{eq:noise-psds}
S_{hh}^{\rm SN}(\nu) &= \frac{\xi_0^2}{G^2 \kappa  m^2 D^2 \nu^4  |\chi_c|^2 |\chi_m|^2} \\
S_{hh}^{\rm BA}(\nu) &= \frac{ G^2 \kappa}{\xi_0^2 m^2 D^2 \nu^4}  \left| \chi_c(\nu) \right|^{2}~,
\end{split}
\end{align}
where the approximation holds for high frequencies. From these expressions, we see that variation of the (drive-enhanced) coupling $G$ trades between the shot and back-action noise. In principle, we can always select a target frequency $\omega_*$ and tune the drive strength so that $G = G_*(\omega_*)$ minimizes the sum of these two quantum noise contributions. The resulting noise level is called the Standard Quantum Limit (SQL) at frequency $\omega_*$. The required coupling strength is
\begin{align}
\begin{split}
\label{eq:Gstar}
G_* & = \frac{\xi_0}{\sqrt{\kappa |\chi_m(\omega_*)|} |\chi_c(\omega_*)| } \\ 
& \approx 20~{\rm MHz}\times \left( \frac{\omega_*/2\pi}{100~{\rm kHz}} \right)^{3/2} \left( \frac{100~{\rm Hz}}{\omega_0/2\pi} \right)^{1/2},
\end{split}
\end{align}
where we are taking the high-frequency limit $\omega_* \gtrsim \kappa \gg \omega_0, \gamma$. Note that the required coupling is independent of the test mass $m$. A central question for us is how close a realistic system can get to achieving this requirement.

As discussed in the introduction, magnetic systems like ours allow for relatively large values of the single-photon coupling \eqref{eq:G0}, and can potentially enable SQL-level measurements at frequencies higher than those achievable with optical readout. To study this quantitatively, we first discuss some basic operational restrictions. The magnetic field gradient $b_z$ is limited by the critical field of the superconducting sphere as $b_z \leq \frac{4}{5} B_c / R$ \cite{Hofer_2019}. We will saturate this inequality and assume that the sphere is lead coated with TiN to maximize its density. Thin TiN films have been shown to achieve critical fields up to $B_c \sim 5~{\rm T}$~\cite{PhysRevB.86.184503}. Under this assumption, the achievable single-photon couplings are of order
\begin{align}
\begin{split}
\label{eq:G0-numbers}
G_0 & = \xi_0 \lambda_{\Phi} \beta B_c R \frac{d\omega_a}{d\Phi} \\
& \approx 0.2~{\rm MHz} \times \left( \frac{B_c}{1~{\rm T}} \right)^{1/2} \left( \frac{\lambda_{\Phi}}{0.1} \right) \left( \frac{\beta}{1} \right),
\end{split}
\end{align}
where we take the circuit's frequency response to an input flux $d \omega_a/d\Phi \approx 2 \pi \times 1~{\rm GHz}$~\cite{Schmidt2024} (see the Supplemental Material for further discussion of these parameters). Comparing Eqs. \eqref{eq:Gstar} and \eqref{eq:G0-numbers}, and recalling that the drive-enhanced coupling is $G = \sqrt{\overline{n}} G_0$, the basic conclusion is that SQL-level noise is achievable in this parameter regime, with moderate amounts of circulating power in the circuit $\overline{n} \lesssim 10^{2}$.

Achieving the SQL at larger mass or higher frequency becomes more difficult. The limit at higher frequency comes from the $G_* \sim \omega_*^{3/2}$ scaling in Eq. \eqref{eq:Gstar}. The scaling with sensor size is more subtle. We can parameterize $m = 4 \pi \rho R^3/3$ and $\omega_0^2 = 24 B_c^2/25 \rho R^2$, using Eq.~\eqref{eq:omega0} and assuming $b_z = 4 B_c/5 R$ as discussed above. In terms of the sphere size $R$, we have $G_* \sim \omega_0^{-1/2} \sim R^{1/2}$, while $G_0 \sim (m \omega_0)^{-1/2} R^{1} \sim R^0$, independent of $R$. Thus larger spheres require stronger driven couplings, or equivalently, larger circulating power $\overline{n}$ in the circuit. Beyond $\overline{n} \approx 10^{2}$, the circuits tend to become non-linear.

Additionally, it may be possible to increase the transduction coefficient $\lambda_\Phi$, relating the flux through the loop to the flux through the SQUID. To achieve this, one could increase the SQUID inductance while maintaining a sufficiently large critical current to remain superconducting, which would require operating in the regime $L_s > L_J$. Furthermore, the circuit frequency $\omega_a$ should be kept fixed, which requires small capacitances, perhaps possible through use of superinductors  ~\cite{peruzzo2020surpassing, peruzzo2021geometric}. Alternatively, one could thread the flux directly into the circuit via $L_a$ rather than $L_s$. Given these options, we also illustrate in Fig. \ref{fig:setup} the reach for which $\lambda_\Phi = 1$, while emphasizing that this would require non-standard SQUID design.

In Fig. \ref{fig:noise-profiles}, we show all the contributions to the noise budget assuming that the SQL is achieved at $\omega_* \approx \kappa/2 \approx 2\pi\times 100~{\rm kHz}$. Tuning the loss rate in this way minimizes the coupling required to reach the SQL at this frequency. We see that the shot noise and backaction cross at the appropriate frequency, and the thermal noise, modeled using the standard formula \cite{kubo1966fluctuation,clerkIntroductionQuantumNoise2010},
\begin{align}
S^{\rm T}_{hh}(\nu) = \frac{2 \gamma T}{m D^2 \nu^4},
\label{eqn:thermalPSD}
\end{align}
is subdominant at the relevant frequencies for $\gamma \approx 10^{-10}$~Hz~\cite{higginsMaglev2023} and $T = 10~\text{mK}$. The same noise budget was used to produce the basic sensitivity projections in Fig. \ref{fig:setup}.

There are additional technical sources of noise, although these can be effectively mitigated. Firstly, external vibrations lead to imprecision noise, which can be reduced by appropriate isolation~\cite{Aldcroft:1992im}. In existing superconducting, magnetomechanical systems~\cite{hofer2022highq}, the vibrational noise at frequencies $ f \gtrsim \SI{30}{Hz}$ has been measured to be $S_{hh}^{1/2} \lesssim \SI{5e-10}{Hz^{-1/2}}$, taking a fiducial baseline of 10 cm to convert to strain units. Extrapolating such a three-stage isolation system to higher frequencies, we expect $S_{hh}^{1/2} \lesssim \SI{4e-25}{Hz^{-1/2}}$ for $f \gtrsim \SI{10}{kHz}$, well below the measurement noise. Lastly, noise in the background magnetic field due to current fluctuations in the primary coils is mitigated by using a persistent superconducting current~\cite{hofer2022highq,1643122,kamerlingh}. A first order low-pass filter with a corner frequency $\lesssim 10~\mathrm{nHz}$ will mitigate this noise below $S_{hh}^{1/2}(6~\mathrm{kHz}) \lesssim 10^{-21}~\text{Hz}^{-1/2}$~\cite{vanDyck1999,hofer2022highq}, below the quantum measurement noise.

\textit{Outlook}
\label{sec:discussion}--- The sensitivity curves shown in Fig. \ref{fig:setup} suggest that levitated superconducting test masses, read out with flux tunable microwave circuits, could be capable of broadband detection of gravitational waves in the 1 kHz--10 MHz band. Above this frequency range, our noise estimates are on less clear footing, because internal mechanical modes in the system will come into play. However, within the main band of interest, our proposal would open a swath of previously unobservable GW frequencies and strains.

To contextualize the potential of such a detector, a 30~kg SLedDoG with $\lambda_{\Phi} = 1$ and a run time of $10^7$~s would be sensitive to monochromatic, coherent GWs with $h \sim 10^{-24}$ at $\sim 10$~kHz - MHz frequencies. Such a sensitivity could, for instance, probe gravitational radiation from superradiant axion clouds around stellar-mass black holes in the Milky Way~\cite{sprague2024simulatinggalacticpopulationaxion} and inspirals of primordial black holes with masses $10^{-12}~M_{\odot} \lesssim m_{\rm PBH} \lesssim 10^{-7}~M_{\odot} $~\cite{Franciolini_2022}. We note, however, that such statements relating strain sensitivity to BSM physics are highly model-dependent. The most directly comparable broadband detection proposal in this frequency band is MAGO~2.0~\cite{Berlin:2023grv} which, for a run time of  $10^7$~s, has sensitivity to coherent, monochromatic GWs down to $h \sim 10^{-21}$ for kHz - MHz frequencies.

Importantly, the core technology is being actively pursued to perform searches for a variety of dark matter candidates \cite{higgins2023maglev} and tests of low-energy quantum gravity \cite{Carney:2018ofe}. The advantage of these magnetic systems is the possibility of strong single-photon couplings, enabling operation at or below the SQL at frequencies well above the kHz scale, a task challenging with standard optical readout methods. Our results further reinforce the need to develop such systems as future quantum-limited detectors of high-frequency signals.

\textit{Acknowledgements}--- We thank Daniel Belkin, John Groh, Roni Harnik, Yoni Kahn, Noah Kurinsky, Achintya Paradkar, Nick Rodd, Philip Schmidt, Jan Sch{\"u}tte-Engel, Natanael Bort Soldevila, Jacob Taylor, Witlef Wieczorek, and Martin Zemlicka for helpful discussions. This material is based upon work supported by the National Science Foundation Graduate Research Fellowship Program under Grant No. DGE 21-46756; by the U.S. DOE, Office of High Energy Physics, under Contract No. DEAC02-05CH11231, the Quantum Information Science Enabled Discovery (QuantISED) for High Energy Physics grant KA2401032, and the Quantum Horizons: QIS Research and Innovation for Nuclear Science Award DE-SC0023672. This research was funded in whole or in part by the Austrian Science Fund (FWF) [10.55776/ESP525].

\bibliography{bib}{}

\hfill \break

\onecolumngrid

\begin{center}
\large{\textbf{End Matter}}
\end{center}

\twocolumngrid
\appendix

\renewcommand{\theequation}{A\arabic{equation}}
\setcounter{equation}{0}

\textit{Appendix A: Total signal calculation}\label{app:complete-signal}--- The signal in Eq. \eqref{eqn:signal} accounts only for the relative motion of the pickup loop and the sphere. At frequencies $>$~kHz, the setup is effectively in free fall. Thus, a GW then affects the following: the sphere–loop separation $\xi$, affecting the flux through the loop; the path of the loop $\mathcal{C}_{\ell}$, upon which the flux depends; distortion of the primary coils, including the shape and separation from the trap center; and the shape of the sphere, which changes the field it induces.

In the proper detector frame, Newtonian forces act on every element of the experimental apparatus due to a passing gravitational wave \cite{Maggiore}. The components of the force-density $f_i$ acting on an object of density $\rho$ at location $\mathbf{x}$ are $f_i = \frac{1}{2} \rho h_{ij} x^j$, where $h_{ij}$ are the components of the metric perturbation in TT-gauge. This force will act on each component of the experimental apparatus, inducing corresponding changes in the flux through the resonator cavity. Restricting our analysis to a $+$-polarized GW, we find a parametric description of the primary coil wires
\begin{align}
    \mathbf{c}_\pm(\lambda) \rightarrow \begin{pmatrix}
        R_C\cos \lambda  \left(1+ h_+(t) \cos^2\theta_h\right) \mp \frac{h_+(t) a}{2} \sin 2\theta_h \\ R_C \sin \lambda \left( 1 - h_+(t) \right) \\ \pm a + h_+(t) \left(\pm a \sin^2\theta_h  - \frac{R_C}{2} \cos \lambda \sin 2 \theta_h  \right)
    \end{pmatrix}.
    \label{eqn:coilPlus-app}
\end{align}
The pick-up loop is simply two circular wires in a gradiometric configuration, meaning it will be distorted just as Eq. \eqref{eqn:coilPlus-app}, with $R_C \to R_\ell$ and $a \to D$. We neglect distortions of the pick-up loop, since the field is concentrated at the center, implying the flux is insensitive to small changes in the loop boundary.

In a gradiometric configuration, one measures the difference between fluxes through the two loops of the gradiometric coil. Thus any change in magnetic field that is translation-invariant in the plane of the loop will not cause a signal. Since both the pick-up loop and the primary coils are rotated by a GW by the same angle, and the quadrupolar field generated by elliptical coils is translation-invariant in directions perpendicular to the primary coil axis, distortion of the primary coils does not induce any change in flux through the gradiometric pick-up loop.

While the displacement of the primary coil and pickup loop cause only a negligible change in flux, changes in the boundary of the primary coils affect the $z$-component of the magnetic field. For a set of elliptical anti-Helmholtz coils with axis lengths $R_p ( 1 + h_x)$ and $R_p (1 + h_y)$ and co-axial distance to the trap-centre $a(1 + h_z)$, the $z$-component of the magnetic field is 
\begin{align}
    \nonumber
    B_z = &\frac{3 I a R_p^2}{(a^2 + R_p^2)^{7/2}} \Big[ a^2(1 + h_x + h_y - 4 h_z) \\
    &+ R_p^2 \left(1 - \frac{3}{2}(h_x + h_y) + h_z\right)\Big] z
\end{align}
to linear order in $h_{x,y,z}$. We note that for a purely $+$-polarised GW travelling along the equator with $\theta_h = \pi/2$, $h_x = 0$, $h_y = - h_+$ and $h_z = h_+$, and so $B_z$ is constant to linear order in $h$ for $a = R_p/\sqrt{2}$. In this configuration, the GW-induced strain of the primary coils may be ignored.

Finally, the sphere itself feels a strain under the influence of a GW. The strain distorts the shape of the sphere, which in turn affects the magnetic field produced by the sphere. It can be shown that GWs couple only to the quadrupolar mechanical modes. Quadrupolar mechanical deformations at $\order{h}$ induce $\order{h}$ monopole, quadrupole, and octopole corrections to the magnetic field sourced by the sphere. We calculate the corrections to the induced flux through a gradiometric pickup loop $\delta \Phi^{(1)}$ for a $+$-polarized GW propagating in the $xy$ plane and find $\delta \Phi^{(1)} / \Phi^{(0)} =  -0.54 h e^{-i \omega_g t}$. We neglect this effect in the main text as it is $\approx 5$ times smaller than the signal coming from variation in the sphere--loop separation.

\renewcommand{\theequation}{B\arabic{equation}}
\setcounter{equation}{0}

\textit{Appendix B: System Hamiltonian and input-output formalism}\label{app:quantization}--- In this appendix, we derive the full Hamiltonian of the sphere coupled to the readout circuit and calculate the signal and noise power spectral densities. Throughout, we follow Ref.~\cite{kuenstner2022quantum}, generalizing where necessary. We begin by deriving the Hamiltonian for the sphere, the readout circuit, and the coupling. Independently, each of the two systems is simply a harmonic oscillator. We write
\begin{align}
    \nonumber
    &H_{\rm sphere} = \frac{p^2}{2m}~ + \frac{1}{2} m \omega_0^2 \xi^2, \\
    \label{eq:H_scs-app}
      &H_{\text{LC}} = \frac{q^2 }{2C}  + \frac{\phi^2}{2(L_a + L_J(\Phi))} = \frac{q^2}{2 C} + \frac{1}{2} C \tilde{\omega}^2_a \phi^2,
\end{align}
where we have restricted the motion of the sphere to the $z$ axis, $\Phi$ is the magnetic flux through the readout circuit, $\tilde{\omega}_a = \tilde{\omega}_a(\Phi) = 1/\sqrt{L(\Phi) C}$ is the (flux-dependent) frequency of the circuit in terms of the total (flux-dependent) inductance $\tilde{L} = \tilde{L}(\Phi) = L_a + L_{J}(\Phi)$, $q$ is the charge on the capacitor, and $\phi$ is the phase. 

Motion of the sphere relative to the pickup loop (with distance $\xi$) induces an external flux $\Phi \approx \eta \xi$ through the resonator circuit, where
\begin{equation}
\label{eq:eta-app}
\eta \equiv \frac{\partial \Phi}{\partial \xi} = \beta b_z R^2 \lambda_{\Phi} \approx \beta b_z \frac{M}{2 L_p} R^2.
\end{equation}
Here, $\beta \sim \order{1}$ is a dimensionless geometric coefficient defined through $\partial \Phi_p / \partial z = \beta R^2 b_z$, $\lambda_{\Phi}$ is the coupling of the flux through the pickup loop and the flux through the resonator circuit, and we have assumed that the input inductance $L_I$ is equal to the pickup inductance so as to maximize the coupling $\lambda_{\Phi}$. We numerically evaluate $\beta$ for a square, gradiometric, pick-up loop coupled to a superconducting sphere and find that at a pick-up loop–sphere separation D equal to R, we achieve a maximum of $\beta \approx 1.6$ for loop of size linear size $\approx 1.1R$.

For small fluxes $\Phi \approx 0$ through the SQUID, in particular those generated by small motions of the sphere, we can expand
\begin{equation}
\label{eq:omega_a-app}
\tilde{\omega}_a(\Phi) = \omega_a + \Phi \frac{\partial \tilde{\omega}_a(0)}{\partial \Phi}, \ \ \ \frac{\partial \tilde{\omega}_a(0)}{\partial \Phi} = - \frac{1}{2} C \omega_a^3 \frac{\partial L_J(0)}{\partial \Phi},
\end{equation}
where here and in the rest of the paper, $\omega_a = 1/\sqrt{L C}$ is the (flux-independent) LC frequency in terms of the (flux-independent) total inductance at zero flux $L = \tilde{L}(0) = L_a + L_J(0)$. Finally, we can use this expansion to write the total Hamiltonian as two oscillators with a simple coupling:
\begin{align}
    \nonumber
    H_{\text{sys}} &=  \frac{p^2}{2m} + \frac{1}{2} m \omega_0^2 \xi^2 + \frac{q^2}{2 C} + \frac{1}{2} C \omega_a^2 \phi^2 + V_{\rm int} \\
    \label{eq:H_coupled_end_matter}
    &=  \frac{p^2}{2m} + \frac{1}{2} m \omega_0^2 \xi^2 + \frac{q^2}{2 C} + \frac{1}{2} C \omega_a^2 \phi^2 - C \omega_a \frac{\partial \omega_a}{\partial \Phi} \eta \xi \phi^2.
\end{align}
The interaction term comes from expanding $\tilde{\omega}_a^2$ and applying Eqs.~(\ref{eq:eta-app} --\ref{eq:omega_a-app}).

This system can be quantized following standard methods, promoting $\xi$, $p$, $\phi$, and $q$ to operators~\cite{blais2021circuit}. The Heisenberg equations of motion for the system are

\begin{align}
\nonumber
&\dot{\xi} = \frac{p}{m}~, \quad\quad \dot{p} = - m \omega_0^2 \xi - \frac{G_0}{\phi_0^2 \xi_0} \phi^2~, \\
\label{eq:EOM_nonlinear-app}
&\dot{\phi} = \frac{q}{C}~, \quad\quad \dot{q} = - C \omega_a^2 \phi - \frac{2 G_0}{\phi_0^2 \xi_0} \phi \xi.
\end{align}
where $G_0$, given in Eq. \eqref{eq:G0-numbers} is the single photon coupling and has units of frequency as usual.
Eqs.~\eqref{eq:EOM_nonlinear-app} describe the sphere and LC circuit in the absence of any noise and without the microwave drive/readout. To incorporate these effects, we use standard input-output techniques~\cite{beckey2023quantum}. The microwave line is assigned input and output fields $\phi_{\rm in}, \ q_{\rm in}$, with effective coupling rate $\kappa$, and we also allow for an external force $F_{\rm in}$ on the mechanical motion of the sphere, which can include both the signal of interest as well as thermal noise. The output fields are related to the input fields by the usual I/O relations
$\phi_{\rm out} = \phi_{\rm in} - \sqrt{\kappa} \phi~, \ q_{\rm out} = q_{\rm in} - \sqrt{\kappa} q~.
$

Driving the microwave line at the LC frequency $\phi_{\rm in} \to \overline{\phi}_{\rm in} \cos (\omega_a t) + \phi_{\rm in}$, where the overlined term is the drive strength and the second term is the small vacuum fluctuation around this drive, we solve for the steady-state solution $\overline{\phi} = \overline{\phi}_{\rm in}/\sqrt{\kappa}$ to leading order in couplings and perturbations, assuming a sufficiently strong drive $|\overline{\phi}_{\rm in}| \gg \phi_{\rm in}$. Moving to the frame co-rotating with the drive (i.e., the LC circuit) and linearizing around the strong drive, we obtain the equations of motion \eqref{eq:EOM_main} given in the main text.

The observable we are interested in is the output charge $q_{\rm out}$, since $q$ is the variable that gets the information about the mechanical system.
In the frequency domain, we obtain the solution for $q_{\rm out}(\nu)$ :
\begin{align}
\nonumber
q_{\rm out}(\nu) =  &e^{i \phi_c(\nu)} q_{\rm in}(\nu) + 2 \left( \frac{G}{\phi_0 \xi_0} \right)^2 \kappa \chi_c^2(\nu) \chi_m(\nu) \phi_{\rm in}(\nu) \\
\label{eq:q_out-app}
&- 2 \left( \frac{G}{\phi_0 \xi_0} \right) \sqrt{\kappa} \chi_c(\nu) \chi_m(\nu) F_{\rm in}(\nu),
\end{align}
where now we use response functions around Eq. \eqref{eq:cavity-response-function}.

Equation (\ref{eq:q_out-app}) shows how both any signals of interest and a variety of noise effects are encoded onto the measured output. The signal is part of $F_{\rm in}$; for a gravitational wave it is $F_{\rm in}^{\rm sig}(\nu) = m \nu^2 D h(\nu)$. Thermal noise acting on the sphere motion will also be part of $F_{\rm in}$ and couple at order $G$. To estimate the strain from the charge data stream we divide by
\begin{equation}
h_E(\nu) = \frac{1}{\frac{2 G}{\phi_0 \xi_0} \sqrt{\kappa} \chi_c(\nu) \chi_m(\nu) m \nu^2 D} q_{\rm out}(\nu).
\end{equation}
The noise power spectrum referred to this observable is then obtained by the Weiner-Khinchin theorem, which amounts to squaring Eq. \eqref{eq:q_out-app} and taking the expectation value:
\begin{align}
\nonumber
S_{hh} = &\frac{1}{\left( \frac{ 2G}{\phi_0 \xi_0} \right)^2 \kappa |\chi_c|^2 |\chi_m|^2 m^2 \nu^4  D^2} S_{qq} \\
&+ \left( \frac{G}{\phi_0 \xi_0} \right)^2 \frac{\kappa |\chi_c|^2}{m^2 \nu^4 D^2} S_{\phi\phi} + \frac{1}{m^2 \nu^4 D^2} S_{FF}.
\end{align}

\newpage




\onecolumngrid

\section{Supplemental Material: Superconducting Levitated 
Detector of Gravitational Waves}
\counterwithin*{equation}{section}
\renewcommand\theequation{\thesection\arabic{equation}}
\renewcommand\thesection{S\arabic{section}}

\begin{center}
Daniel Carney, Gerard Higgins, Giacomo Marocco, Michael Wentzel
\end{center}

\hfill \break

\renewcommand{\theequation}{S\arabic{equation}}
\setcounter{equation}{0}

\let\clearpage\relax
\let\newpage\relax
\onecolumngrid

This supplemental material is organized as follows. First, we perform a complete calculation of the signal and show that the dominant signal is from the center of mass motion of the sphere. We then calculate the signal and noise power spectral densities using the full Hamiltonian of the sphere coupled to the readout circuit. Finally, we present a numerical calculation of the dimensionless coupling $\beta$ and shows that it is $\order{1}$.

\section{Complete Signal Calculation}\label{sec:complete-signal-sup}
The signal considered in the main text only takes into account the relative motion of the pickup loop and the sphere under the influence of a GW. However, at the frequencies we consider here ($>$~kHz), the entire setup is effectively in free fall. This result can be seen by noting that the setup is made of solid materials with sound speeds $v_s \sim 10^3~\textrm{m}/\textrm{s}$ and has a length scale of $1$~m yielding a resonant frequency $\omega_c \sim \textrm{kHz}$. In practice, we must therefore consider the motion of the anti-Helmholtz (AHC) coils that source the quadrupolar field, the motion of the sphere, and the motion of the pickup loop. We will show that the gradiometric (figure-8) pickup circuit setup nullifies any change in primary magnetic field due to the motion of the primary coils. The dominant effect is due to the change in field sourced by the superconducting sphere.

A passing gravitational wave affects the following:
\begin{itemize}
\item The distance to the loop $D$, and thus the sphere--loop separation $\xi$, affecting the flux through the loop. This is the dominant signal we consider, and is discussed in the main body of the text.
\item The path of the loop $\mathcal{C}_\ell$, upon which the flux depends, which we show to be subleading to the previous effect.
\item Distortion of the primary coils, including their shape and separation from the trap center. As we show, in the gradiometric loop configuration which we employ, this does not directly induce a change in flux through the loop, since the magnetic field is still approximately translation-invariant. However, both of these effects can change the magnetic field gradient along the coaxial direction at the centre of the quadrupole trap, which in turn changes the magnetic field that the sphere produces. This field is not translation-invariant, and so produces a measurable change through the pick-up loop. This effect can be of the same order as the first effect, but there exist configurations in which it vanishes.
\item The shape of the sphere, which changes the field it induces. We show that this effect is subdominant. 

\end{itemize}

\subsection{Forces due to a gravitational wave}

In this section, we derive results on how a gravitational wave affects the primary coils, pick-up loop, and levitated particle.

In the proper detector frame, Newtonian forces act on every element of the experimental apparatus due to a passing gravitational wave \cite{Maggiore}. The components of the force-density $f_i$ acting on an object of density $\rho$ at location $\mathbf{x}$ are $f_i = \frac{1}{2} \rho h_{ij}^\mathrm{TT} x^j$, where $h_{ij}^\mathrm{TT}$ are the components of the metric perturbation in TT-gauge. From here on, we omit the superscript TT on the metric perturbation $h$.

For a gravitational wave travelling in the $\hat{z}$ direction, the components of a monochromatic metric perturbation $h'$ of energy $\omega_g$ are
\begin{align}
h'(t,z) = \begin{pmatrix}
    h_+ & h_\times & 0 \\
    h_\times & -h_+ & 0 \\
    0 & 0 & 0
\end{pmatrix} \cos\big(\omega_g(t-z)\big),
\label{eqn:TTmetric}
\end{align}
which depends on the amplitude of the two polarisations $h_+$ and $h_\times$.

Given the cylindrical symmetry of our system, we may consider a gravitational wave to lie in the $(x,z)-$plane, without loss of generality. We obtain such a gravitational wave, propagating with a polar angle $\theta_h$ from the $x-$axis, by rotating Eq. \eqref{eqn:TTmetric} $h(\theta_h) = \mathcal{R}(\theta_h) \cdot h' \cdot \mathcal{R}^\top(\theta_h)$, with
\begin{align}
    \mathcal{R}(\theta_h) = 
    \begin{pmatrix}
        \cos \theta_h & 0 & \sin \theta_h \\
        0 & 1 & 0 \\
        -\sin\theta_h & 0 & \cos\theta_h
    \end{pmatrix}.
\end{align}
The force density due to such a gravitational wave may be decomposed into its two polarisation components $\mathbf{f}_+$ and $\mathbf{f}_\times$, with
\begin{align}
    \mathbf{f}_+(\mathbf{x}) = h_+ \begin{pmatrix}
        x \cos^2 \theta_h - \frac{z}{2}\sin 2 \theta_h \\
        - y \\
       z \sin^2\theta_h - \frac{x}{2}\sin 2\theta_h 
    \end{pmatrix} \cos\omega_g t,
    \label{eqn:fPlus}
\end{align}
and
\begin{align}
    \mathbf{f}_\times(\mathbf{x}) = h_\times \begin{pmatrix}
        y \cos\theta_h \\
        x \cos \theta_h - z \sin\theta_h \\
        -y \sin\theta_h
    \end{pmatrix} \cos\omega_g t,
    \label{eqn:fCross}
\end{align}
where we have assumed that $\omega$ is much smaller than the inverse size of the experiment.

\subsubsection{Force on primary coils}
The primary coils comprising the anti-Helmholtz system, in the absence of a gravitational wave, are two circular wires of radius $R_C$ at heights $z = \pm a$; a parametric description $\mathbf{c}_\pm(\lambda)$ of these two undisturbed coils is
\begin{align}
    \mathbf{c}_\pm(\lambda) = \begin{pmatrix}
        R_C \cos \lambda \\ R_C \sin \lambda \\ \pm a
    \end{pmatrix},
\end{align}
where $\lambda \in [0, 2\pi)$.

Under the effect of a $+$-polarised GW, we approximate the primary coils as a set of test masses, such that they are distorted by the force Eq. \eqref{eqn:fPlus} a
\begin{align}
    \mathbf{c}_\pm(\lambda) \rightarrow \begin{pmatrix}
        R_C\cos \lambda  \left(1+ h_+ \cos^2\theta_h\right) \mp \frac{h_+ a}{2} \sin 2\theta_h \\ R_C \sin \lambda \left( 1 - h_+ \right) \\ \pm a + h_+ \left(\pm a \sin^2\theta_h  - \frac{R_C}{2} \cos \lambda \sin 2 \theta_h  \right)
    \end{pmatrix},
    \label{eqn:coilPlus}
\end{align}
where we have absorbed the time-dependence of the GW into $h_+$.
This equation describes an elliptical coil that is subject to five effects: i) the length along the $x$-axis is changed to $R_C(1 + h_+ \cos^2 \theta_h)$; (ii) the length along the $y$-axis is changed to $R_C (1 - h_+ )$, (iii) the coil is $x$-translated by $\mp \frac{h_+ a}{2} \sin2\theta_h $; (iv) the coil is $z$-translated by $\pm a h_+ \sin^2 \theta_h$; and (v) the coil is rotated through the $y$-axis by an angle $\phi_h \approx -\frac{h_+}{2} \sin 2 \theta_h$. 

Similarly, for a $\times$-polarised wave, the coils are distorted by Eq. \eqref{eqn:fCross} resulting in
\begin{align}
    \mathbf{c}_\pm(\lambda) \rightarrow \begin{pmatrix}
        R_C \left( \cos \lambda + h_\times \sin \lambda \cos \theta_h \right) \\
        R_C \left( \sin \lambda + h_\times \cos\lambda \cos\theta_h \right) \mp a h_\times \sin\theta_h \\
        \pm a - h_\times R_C \sin \lambda \sin \theta_h,
    \end{pmatrix}
    \label{eqn:coilCross}
\end{align}
which again describes an ellipse, but now i) with axis lengths $R_C( 1 \pm h_\times \cos \theta_h)$; (ii) $y$-translated by $\mp a h_\times \sin\theta_h$; and (iii) tilted around the $x$-axis by $\phi_h \approx -h_\times \sin \theta_h$.

\subsubsection{Force on pick-up loop}
\label{sec:pul}
The pick-up loop is comprised of a two circular lengths of wire in a figure-8 shape. As such, this loop will be distorted in an analogous manner to Eqs. \eqref{eqn:coilPlus} and \eqref{eqn:coilCross}, with the replacement $R_C \to R_\ell$ and $a \to D$.

We neglect distortions of the pick-up loop, since, as shown in Fig. \ref{fig:Bz}, the field is concentrated at the center of the loop, implying the flux is insensitive to small changes in the loop boundary.

In a gradiometric configuration, one measures the difference between fluxes through the two loops of the figure-8 coil. Thus any change in magnetic field that is translation-invariant in the plane of the loop will not cause a signal. Since both the pick-up loop and the primary coils are rotated by a GW by the same angle, and the quadrupolar field generated by elliptical coils is still translation-invariant in directions perpendicular to the primary coil's axis, the distortion of the primary coils does not induce any change in flux through the gradiometric pick-up loop.  

\subsection{Magnetic fields sourced by Anti-Helmoltz Coils}
\label{sec:primaries}

The quadrupolar trap is sourced by the magnetic field of two on axis solenoids with current running in opposite directions. We will assume that without a GW, the solenoids are fixed at $a_{\pm} \hat{z}$. We will refer to quantities relating to the solenoid at $a_+\hat{z}$ with the subscript $+$ and those relating to the solenoid at $a_-\hat{z}$ with the subscript $-$. We'll start by solving for the magnetic field sourced by a generic current loop of radius $R_{\ell}$ and height $a$ positioned in the $xy$ plane. The magnetic field can be written in integral form as
\begin{equation}
    \label{eq:loop-b-field-integral}
    \mathbf{B}(r, \theta, z) = \frac{I R_{\ell}}{4 \pi} \int_0^{2 \pi} \frac{d\theta_{\ell} \hat{\boldsymbol{\theta}} \times (\mathbf{r}_{\ell} - \mathbf{r}) }{ |\mathbf{r}_{\ell} - \mathbf{r}|^{3}}
\end{equation}
In cylindrical coordinates, the separation vector between a point on the loop (subscripts $\ell$) and a point in space is
\begin{equation}
    \label{eq:sep-vector}
    \mathbf{r}_{\ell} - \mathbf{r} = \left( r \cos(\theta_{\ell} - \theta) - R_{\ell} \right) \hat{r}  + r \sin (\theta - \theta_{\ell}) \hat{\theta} + (z - a) \hat{z}
\end{equation}
Therefore, the magnetic field is
\begin{equation}
    \label{eq:loop-b-field-integral-2}
    \mathbf{B}(r, \theta, z) = \frac{\mu_0 I R_{\ell}}{4 \pi} \int_0^{2 \pi} d\theta_{\ell} \frac{- ( r \cos(\theta - \theta_{\ell}) - R_{\ell}) \hat{z} + (z-a)\hat{r}}{\left( r^2 + R_{\ell}^2 - 2R_{\ell}r \cos(\theta - \theta_{\ell}) + (z-a)^2 \right)^{3/2}}
\end{equation}

We consider a system in which the radii of both the sphere and the pickup loop as well as the loop-sphere displacement are small compared to the radius and the z position of the primary coils. In this case, we can expand the integrand and solve for the magnetic field along the $z$-axis -- this is the only component of the field that affects the sphere-induced field as well as the flux through pick-up loops in the $x-y$ plane.
\begin{equation}
    \label{eq:Bz-expansion}
    B^{\rm 1 coil}_z(z) = \frac{\mu_0 I R_{\ell}^2}{2} \frac{a^2 + R_{\ell}^2 +3 a z}{(a^2 + R_{\ell}^2)^{5/2}}
\end{equation}
The total field at a point z due to an AHC is
\begin{equation}
    \label{eq:Bz-tot-expansion}
    B^{\rm AHC}_z(z) = \mu_0 I R_{\ell}^2 \frac{3 a z}{(a^2 + R_{\ell}^2)^{5/2}} = b_z z
\end{equation}
where we have defined the magnetic field gradient near the center of the trap as
\begin{equation}
    \label{eq:bz}
    b_z \equiv \mu_0 I R_{\ell}^2 \frac{3 a}{(a^2 + R_{\ell}^2)^{5/2}}.
\end{equation}
Note that Eq. \eqref{eq:Bz-tot-expansion} is translation-invariant along both the $x$ and $y$ axes, and so the flux due to the primary coils through a gradiometric pick-up loop vanishes.

Since the gradient of the magnetic field does vary with the major/minor axes of the ellipse as well as with the separation $a$ of the primary coils, the field that the sphere produces is affected by the motion of the primary coils, which induces a change in flux through the gradiometric loop since it is not translation-invariant.

\subsubsection{Distortions}

For a set of elliptical anti-Helmholtz coils whose axes have length $R_p ( 1 + h_x)$ and $R_p (1 + h_y)$, and whose co-axial distance to the trap-centre is $a(1 + h_z)$, the $z$-component of the magnetic field is 
\begin{align}
    B_z = \frac{3 I a R_p^2}{(a^2 + R_p^2)^{7/2}} \Big[ a^2(1 + h_x + h_y - 4 h_z) + R_p^2 \left(1 - \frac{3}{2}(h_x + h_y) + h_z\right)\Big] z
\end{align}
to linear order in $h_{x,y,z}$.
We note that for a purely $+$-polarised GW travelling along the equator with $\theta_h = \pi/2$, $h_x = 0$, $h_y = - h_+$ and $h_z = h_+$, and so $B_z$ is constant to linear order in $h$ for $a = R_p/\sqrt{2}$. In this configuration, the GW-induced strain of the primary coils may be ignored.

\begin{figure}[htp]
\centering
\includegraphics[width=.48\textwidth]{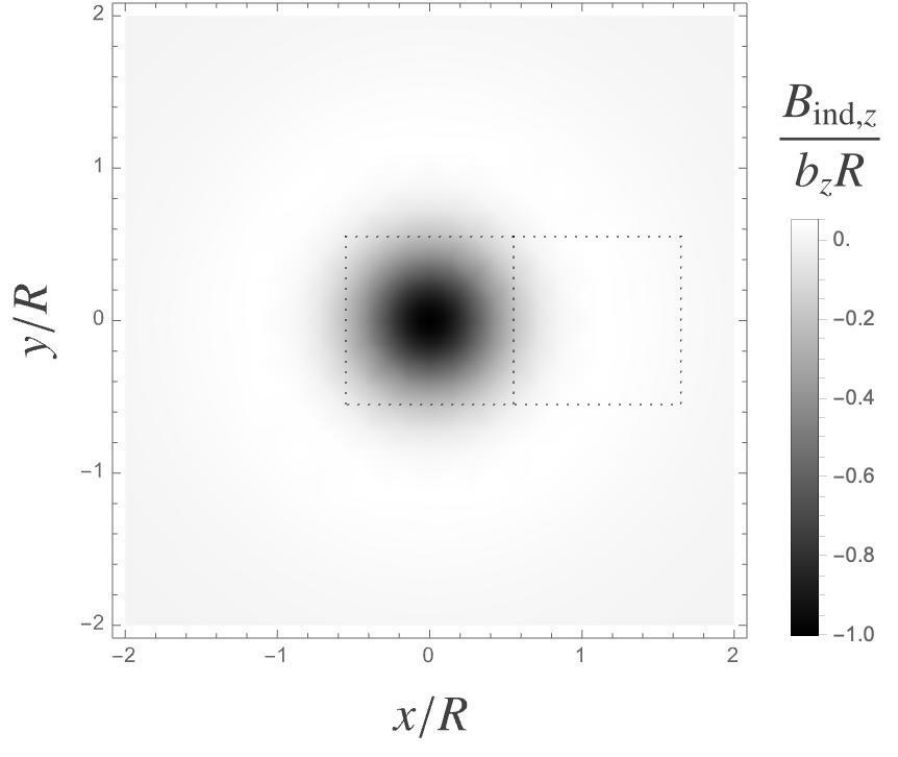}
\caption{The $z$-component magnetic field induced by the sphere, evaluated at the height of the pick-up loop, in units of $b_z \, R$. The pick-up loop is shown in dotted black.}
\label{fig:Bz}
\end{figure}

\subsection{Strain on sphere}
\label{sec:sphere}

The sphere itself feels a strain under the influence of a GW. This strain distorts the shape of the sphere, which in turn affects the magnetic field produced by the currents circulating along its surface. In the quasistatic limit, the induced magnetic field $\mathbf{B}_\mathrm{ind}$ satisfies both $\nabla \cdot \mathbf{B}_\mathrm{ind} = \nabla \times \mathbf{B}_\mathrm{ind} = 0$, and so admits the expansion \cite{higgins2023maglev}
\begin{align}
    \mathbf{B}_\mathrm{ind}(r,\theta,\phi) = \sum_{l=0}^\infty \sum_{m=-l}^l a_{lm} r^{-l-2} \Big[ -(l+1) \mathbf{Y}_{lm} + \boldsymbol{\Psi}_{lm}\Big],
    \label{eqn:inducedFieldExpansion}
\end{align}
where the basis vectors are given in terms of the real scalar spherical harmonics $Y_{lm}(\theta, \phi)$ as
\begin{align}
    \mathbf{Y}_{lm} = Y_{lm} \hat{r}, \qquad \boldsymbol{\Psi}_{lm} = r \nabla Y_{lm}.
\end{align}

The total magnetic field $\mathbf{B}_\mathrm{tot}$ must satisfy the continuity condition  \begin{align}\hat{n} \cdot (\mathbf{B}_0 + \mathbf{B}_\mathrm{ind}) \vert_\Sigma = 0
,\end{align} where $\hat{n} $ is the unit normal to the superconducting element's surface $\Sigma$, and $\mathbf{B}_0$ is the trap magnetic field. This continuity condition specifies the value of the coefficients $a_{lm}$ of Eq. \eqref{eqn:inducedFieldExpansion}, given a particular trap field.

The field due to the sphere, when unperturbed by a GW, satisfies $\hat{r} \cdot \mathbf{B}_\mathrm{tot}(\mathbf{r}_0) = 0$, for all $\mathbf{r}_0 \in \Sigma_0$, where $\Sigma_0$ is surface of the unperturbed sphere. If the sphere is at the centre of a \textit{quadrupolar} trap, we have
\begin{align}
    \mathbf{B}_0 = \sqrt{\frac{\pi}{5}} b_{z} r (2 \mathbf{Y}_{20} + \boldsymbol{\Psi}_{20}),
    \label{eqn:quadrupoleHarmonics}
\end{align}
where $b_z$ is given by Eq. \eqref{eq:bz}. The continuity condition $B_{\mathrm{tot},r}(R,\theta,\phi) = 0$ implies that the unperturbed coefficients $a_{lm} = a^{(0)}_{lm}$ are
\begin{align}a^{(0)}_{20} = \frac{2}{3}\sqrt{\pi/5} \, b_z R^5,
\label{eqn:sphereInducedCoefficient}
\end{align} where $R$ is the sphere's radius, with the coefficients vanishing. The $z$-component of this induced field is plotted in Fig. \ref{fig:Bz}, where we have set $z=R$.

Under  small perturbation due to a GW, the continuity condition implies 
\begin{align}
    \hat{n} \cdot \mathbf{B}_\mathrm{tot} \approx \hat{r} \cdot \mathbf{B}_\mathrm{tot}(\mathbf{r}_0) + \hat{\delta} \cdot \mathbf{B}_\mathrm{tot}(\mathbf{r}_0) + \hat{r}_i \cdot \Delta_j \nabla_j B_{\mathrm{tot},i}(\mathbf{r}_0) + \mathcal{O}(h^2) = 0,
    \label{eqn:continuityExpansion}
\end{align}
where $\hat{\delta} \equiv \hat{n} - \hat{r}$ is the deviation of the perturbed normal vector, and $\boldsymbol{\Delta}(\theta, \phi) \equiv \mathbf{r} - \mathbf{r}_0$ is the change in position of the sphere's surface, both of which are $\mathcal{O}(h)$. We also expand the induced magnetic field in terms of its unperturbed value $\mathbf{B}^{(0)}_\mathrm{ind}$ and the $\mathcal{O}(h)$ corrections $\mathbf{B}^{(1)}_\mathrm{ind}$; Eq. \eqref{eqn:continuityExpansion} then implies
\begin{align}
    \hat{r} \cdot \mathbf{B}^{(1)}_\mathrm{ind} = - \hat{\delta}\cdot \Big( \mathbf{B}_0(\mathbf{r}_0) + \mathbf{B}^{(0)}_\mathrm{ind}(\mathbf{r}_0) \Big)  - \boldsymbol{\Delta}\cdot \nabla \Big( B_{0,r}(\mathbf{r}_0) + B^{(0)}_{\mathrm{ind},r}(\mathbf{r}_0) \Big), 
    \label{eqn:hContinuity}
\end{align}
which must be solved to find the coefficients $a^{(1)}_{lm}$ of the corrections to the induced field.

Assuming that the frequency $\omega_g$ of a GW incident on the sphere is much larger than the mechanical resonance of the sphere, the distortion of the sphere due to the GW can be decomposed into a time dependent amplitude and dimensionless spatial mode profiles $\boldsymbol{S}_{nlm}$:
\begin{equation}
    \boldsymbol{\Delta}(\boldsymbol{x}, t) = h \eta^g_{nlm} e^{-i \omega_g t} R \boldsymbol{S}_{nlm}(\boldsymbol{x})~.
\end{equation}
The amplitude of the mechanical perturbation depends on the GW strain, the sphere radius, and the GW coupling to the mechanical mode $\eta^g_{nlm}$ defined as~\cite{Berlin:2023grv}
\begin{equation}
    \eta^g_{nlm} = \frac{1}{2 V R} \hat{h}^{\rm TT}_{ij} \int_V dV x^j S^i_{nlm}(\boldsymbol{x})~,
\end{equation}
where $\hat{h}^{\rm TT}$ is the TT metric perturbation normalized by the strain. The spatial profiles, normalized such that $\int_V dV |\boldsymbol{S}(\boldsymbol{x})|^2 = V$, can be expanded in terms of vector spherical harmonics such that each mode causes a distortion~\cite{Berlin:2023grv,Maggiore} 
\begin{align}
    \label{eq:sphere-perturbation}
    \boldsymbol{\Delta}_{nlm}(\boldsymbol{x}) = h \eta^g_{nlm} R \left(A_{nl}(R)\mathbf{Y}_{lm} + B_{nl}(R) \boldsymbol{\Psi}_{lm} \right)~,
\end{align}
where the coefficients $A_{nl}, B_{nl}$ depend on the mechanical properties of the sphere. 

The unit normal vector given in terms of the sphere's deformed radius $R_h(\mathbf{r}) := | \mathbf{r_0} - \Delta(\mathbf{r})|$ can be determined from Eq.~\eqref{eq:sphere-perturbation} as 
\begin{align}
\begin{split}
    \hat{n} &= \frac{\nabla R_h(\mathbf{r})}{|\nabla R_h(\mathbf{r})|} \\
    &\approx \hat{r} -  h \eta^g_{nlm} A_{nl} \boldsymbol{\Psi}_{lm},
    \end{split}
\end{align}
which is normal to the perturbed surface at the same $(\theta, \phi)$ as the unperturbed sphere.
Therefore the $\mathcal{O}(h)$ corrections to the induced field must satisfy, by Eq. \eqref{eqn:hContinuity}, 
\begin{align}
\sum_{l'm'}a_{l'm'} \frac{(l'+1)}{r^{l'+2}}Y_{l'm'} = h \eta^g_{nlm} \frac{ A_{nl} \,  a^{(0)}_{20}}{R^4} \left(\boldsymbol{\Psi}_{lm} \cdot \boldsymbol{\Psi}_{20} - 12 \mathbf{Y}_{lm} \cdot \mathbf{Y}_{20}  \right) + \sqrt{\frac{\pi}{5}} h \eta^g_{nlm} A_{nl} b_z R \left( \boldsymbol{\Psi}_{lm} \cdot \boldsymbol{\Psi}_{20} - 2 \mathbf{Y}_{lm} \cdot \mathbf{Y}_{20}\right).\label{eqn:contCorrection}
\end{align}

Now we make use of the facts that spherical harmonics are orthogonal when integrated over the sphere
\begin{align}
    \int d\Omega \, Y_{lm} Y_{l'm'} = \delta_{ll'} \, \delta_{mm'},
\end{align}
and further satisfy the condition
\begin{align}
\begin{split}
    C^{l_1 l_2 l_3}_{m_1 m_2 m_3} &:= \int d\Omega\, Y_{l_1 m_1} Y_{l_2 m_2}Y_{l_3 m_3} \\
    &= \sqrt{\frac{(2l_1 +1)(2l_2 +1)(2l_3 +1) }{4\pi} }\begin{pmatrix}
        l_1 & l_2 & l_3 \\ m_1 & m_2 & m_3
    \end{pmatrix}
    \begin{pmatrix}
        l_1 & l_2 & l_3 \\ 0 & 0 & 0
    \end{pmatrix}, \\
    D^{l_1 l_2 l_3}_{m_1 m_2 m_3} &:= \int d\Omega\, Y_{l_1 m_1} (\nabla_\Omega Y_{l_2 m_2})\cdot \nabla_\Omega(Y_{l_3 m_3}) \\
    &= \frac{l_2(l_2+1)+l_3(l_3+1)-l_1(l_1+1)}{2}\int d\Omega\, Y_{l_1 m_1} Y_{l_2 m_2}Y_{l_3 m_3},
    \end{split}
\end{align}
where $\begin{pmatrix}
        l_1 & l_2 & l_3 \\ m_1 & m_2 & m_3
    \end{pmatrix}$
is the Wigner 3-$j$ symbol, and $\nabla_\Omega$ is the surface gradient. Using these expressions, we may integrate Eq. \eqref{eqn:contCorrection} over the sphere, weighted by a spherical harmonic, to find
    \begin{align}
    \begin{split}
        a^{(1)}_{l'm'} &= - h \eta^g_{nlm} \frac{A_{nl} R^{l'+2}}{l'+1} \Big[\frac{a^{(0)}_{20}}{R^4}\left( 12 C^{l'l2}_{m'm0} - D^{l'l2}_{m'm0}\right)  + \sqrt{\frac{\pi}{5}} b_z R \left( 2 C^{l'l2}_{m'm0} - D^{l'l2}_{m'm0}\right)\Big] \\
        &= - h \eta^g_{nlm} \frac{A_{nl} R^{l'+2}}{l'+1} \frac{a^{(0)}_{20}}{R^4}\left(15 C^{l'l2}_{m'm0} - \frac{5}{2}D^{l'l2}_{m'm0} \right),
        \end{split}
        \label{eqn:pertCoeff}
    \end{align}
    where we have made use of Eq. \eqref{eqn:sphereInducedCoefficient} in the second line.

For the quadrupolar $l=2$ mechanical deformations induced by a GW, the Wigner 3-$j$ symbols appearing in Eq. \eqref{eqn:pertCoeff} vanish unless $l=0,2,4$, implying that $\mathcal{O}(h)$ corrections are induced to the monopole, quadrupole and octupole moments. Furthermore, the GW-mechanical couplings $\eta^g_{nlm}$ are $\order{1}$ only for $n = 0$ and $l = 2$. For a $+$-polarized GW propagating in the $xy$ plane, $\eta^g_{022} = 0.52$ and $\eta^g_{020} = -0.30$ with all other $n=0$ couplings equal to zero; we also find $A_{02} = 0.96$. For such a wave, the change in flux $\delta \Phi^{(1)}$ is then given by integrating the perturbed magnetic field, given by Eqs. \eqref{eqn:inducedFieldExpansion} and \eqref{eqn:pertCoeff}, over the gradiometric loop configuration sketched in Fig. \ref{fig:Bz}, and is
\begin{align}
    \frac{\delta \Phi^{(1)}}{\Phi^{(0)}} = -0.54 h e^{-i \omega_g t};
\end{align}
we neglect this effect in the main text as it is smaller than the signal of Eq. \eqref{eqn:StrainFlux} coming from variation in the sphere--loop separation.

\section{System Hamiltonian and input-output formalism} \label{sec:quantization-sup}

In this appendix, we begin by deriving the full Hamiltonian of the sphere coupled to the readout circuit. We then show how to employ the standard input-output formalism from quantum optics~\cite{beckey2023quantum} to calculate the signal and noise power spectral densities. Throughout, we follow closely Ref.~\cite{kuenstner2022quantum}, generalizing where necessary.

We begin by deriving the Hamiltonian for the sphere, the readout circuit, and their coupling. Independently, each of the two systems is simply a harmonic oscillator. We write
\begin{equation}
    \label{eq:H_scs}
    H_{\rm sphere} = \frac{p^2}{2m}~ + \frac{1}{2} m \omega_0^2 \xi^2, \ \ \  H_{\text{LC}} = \frac{q^2 }{2C}  + \frac{\phi^2}{2(L_a + L_J(\Phi))} = \frac{q^2}{2 C} + \frac{1}{2} C \tilde{\omega}^2_a \phi^2,
\end{equation}
where we have restricted the motion of the sphere to the $z$ axis, $\Phi$ is the magnetic flux through the readout circuit, $\tilde{\omega}_a = \tilde{\omega}_a(\Phi) = 1/\sqrt{L(\Phi) C}$ is the (flux-dependent) frequency of the circuit in terms of the total (flux-dependent) inductance $\tilde{L} = \tilde{L}(\Phi) = L_a + L_{J}(\Phi)$, $q$ is the charge on the capacitor and $\phi$ is the phase. These are conjugate variables which behave analogously to the position and momentum $z$ and $p$ for the mechanical system. The inductance $L_J$ is assumed to be nonlinear, which can be achieved for example using a Josephson interferometer for which $L_J$ is periodic in $\Phi$. 

Motion of the sphere relative to the pickup loop (with distance $\xi$) induces an external flux $\Phi \propto \xi$ through the resonator circuit. For small changes in position, the change in flux is determined by the gradient of the flux with respect to the separation $\partial \Phi_p / \partial z$. We parametrize this gradient as \cite{Schmidt2024}
\begin{align}
    \frac{\partial \Phi_p}{\partial z} = \beta b_z R^2,
\end{align}
where $b_z = \partial (\mathbf{B}_0)_z/\partial z$ is the trap's magnetic field gradient, $R$ is the sphere's radius, and $\beta \sim O(1)$ is a dimensionless geometric coefficient. The magnetic field on the surface of the sphere is approximately $b_z R$, and so this product is constrained to be less than the critical field of the superconducting sphere. The loop flux $\Phi_p$ is linearly related to flux through the SQUID $\Phi = \lambda_\Phi \Phi_p$, through a transduction coefficient
\begin{align}
    \lambda_\Phi = \frac{M}{L_I + L_p + L_w} \approx \frac{M}{2 L_p}
    \label{eqn:fluxTransduction}
\end{align}
where $M \leq \sqrt{L_S L_I}$ is the mutual inductance, given in terms of the SQUID and input inductances $L_S$ and $L_I$, respectively, $L_p$ is the loop inductance, and $L_w$ is any stray inductance in the system. We assume that there is perfect coupling between the input coil and the SQUID, such that $M = \sqrt{L_S L_I}$. The transduction coefficient is maximised for $L_I = L_p$ and $L_w \ll L_I$, which is the approximation in the second step. Integrating $\Phi = \lambda_{\Phi} \Phi_p$ once and putting this together reproduces the result from the main text, which here is more usefully written as
\begin{equation}
\label{eq:eta}
\eta \equiv \frac{\partial \Phi}{\partial \xi} = \beta b_z R^2 \lambda_{\Phi} \approx \beta b_z \frac{M}{2 L_p} R^2.
\end{equation}
For small fluxes $\Phi \approx 0$ through the SQUID, in particular those generated by small motions of the sphere, we can expand
\begin{equation}
\tilde{\omega}_a(\Phi) = \omega_a + \Phi \frac{\partial \tilde{\omega}_a(0)}{\partial \Phi}, \ \ \ \frac{\partial \tilde{\omega}_a(0)}{\partial \Phi} = - \frac{1}{2} C \omega_a^3 \frac{\partial L_J(0)}{\partial \Phi},
\end{equation}
where here and in the rest of the paper, $\omega_a = 1/\sqrt{L C}$ is the (flux-independent) LC frequency in terms of the (flux-independent) total inductance at zero flux $L = \tilde{L}(0) = L_a + L_J(0)$. Finally, we can use this expansion to write the total Hamiltonian as two oscillators with a simple coupling:
\begin{equation}
    \label{eq:H_coupled}
    H_{\text{sys}} =  \frac{p^2}{2m} + \frac{1}{2} m \omega_0^2 \xi^2 + \frac{q^2}{2 C} + \frac{1}{2} C \omega_a^2 \phi^2 + V_{\rm int}.
\end{equation}
The interaction term comes from expanding
\begin{equation}
\tilde{\omega}_a^2 \approx \omega_a^2 + 2 \omega_a \frac{\partial \tilde{\omega}_a}{\partial \xi} \xi \approx \omega_a^2 + 2 \omega_a \frac{\partial \tilde{\omega}_a}{\partial \Phi} \frac{\partial \Phi}{\partial \xi} \xi 
\end{equation}
and using the above results to evaluate the derivatives, yielding
\begin{equation}
\label{eq:Vint}
V_{\rm int} = - C \omega_a \frac{\partial \omega_a}{\partial \Phi} \eta \xi \phi^2,
\end{equation}
where again $\eta$ is given in Eq. \eqref{eq:eta}.

The coupling term in Eq.~\eqref{eq:Vint} represents a nonlinear coupling between the sphere and the LC circuit $\sim \xi \phi^2$. This is similar to the usual optomechanical coupling $\sim x a^\dag a$ between the position of a suspended mirror and the cavity photon number operator. There is a slight difference since the optomechanical coupling is dispersive (position couples to energy, which is conserved) while the magnetomechanical coupling here is not (position couples to $\phi^2$, which is not conserved). However, when the readout system is driven, the two couplings behave identically, as we will see below.

This system can be quantized following standard methods~\cite{blais2021circuit}. Simply, $\xi,p,q$, and $\phi$ are promoted to operators. Since the system is a pair of coupled harmonic oscillators we explicitly introduce ladder operators
\begin{equation}
    \label{eq:scs-ladder}
    \xi = \xi_0 \left(b^{\dagger} + b \right), \ \ p = i p_0 \left(b^{\dagger} - b \right), \ \ \phi = \phi_0 \left(a^{\dagger} + a \right), \ \  q = i q_0 \left(a^{\dagger} - a \right).
\end{equation}
Here the prefactors are the vacuum amplitudes (i.e., the uncertainties in the ground state):
\begin{equation} 
\xi_0 = \sqrt{\frac{1}{2 m \omega_0}}, \ \ p_0 = \sqrt{\frac{ m \omega_0}{2}}, \ \ \phi_0 = \sqrt{\frac{1}{2 C \omega_a}}, \ \ q_0 = \sqrt{\frac{C \omega_a}{2}}.
\end{equation} 
The commutation relations are canonical: $[\xi,p] = [\phi,q] = i$. In terms of the ladder operators we can write the Hamiltonian as
\begin{equation}
    \label{eq:H_ladder}
    H_{\text{sys}} = \omega_a a^{\dagger} a + \omega_0 b^{\dagger} b - G_0
    \left(a^{\dagger} + a \right)^2 \left(b^{\dagger} + b\right)~,
\end{equation}
where the single-photon coupling is
\begin{equation}
G_0 = C \omega_a \eta \frac{\partial \omega_a}{\partial \phi} \xi_0 \phi_0^2 = \frac{1}{2} \eta \xi_0 \frac{\partial \omega_a}{\partial \Phi}, 
\end{equation}
and has units of frequency as usual. The Heisenberg equations of motion are thus
\begin{align}
\begin{split}
\label{eq:EOM_nonlinear}
\dot{\xi} = \frac{p}{m}~, \quad\quad \dot{p} = - m \omega_0^2 \xi - \frac{G_0}{\phi_0^2 \xi_0} \phi^2~, \quad\quad \dot{\phi} = \frac{q}{C}~, \quad\quad \dot{q} = - C \omega_a^2 \phi - \frac{2 G_0}{\phi_0^2 \xi_0} \phi \xi.
\end{split}
\end{align}
We see that the motion of the sphere $\xi$ is imprinted on the circuit, and vice versa, the circuit drives the position of the sphere, through the $G_0$ terms.

Eqs.~\eqref{eq:EOM_nonlinear} describe the sphere and LC circuit in the absence of any noise and without the microwave drive and readout. To incorporate these effects, we use standard input-output techniques~\cite{beckey2023quantum}. The microwave line is assigned input and output fields $\phi_{\rm in},q_{\rm in}$, with effective coupling rate $\kappa$, and we also allow for an external force $F_{\rm in}$ on the mechanical motion of the sphere, which can include both the signal of interest as well as thermal noise. The Heisenberg equations, Eq.~\eqref{eq:EOM_nonlinear}, become Heisenberg-Langevin equations,
\begin{equation}
\begin{aligned}
\label{eq:EOM_nonlinear_2}
\dot{\xi} & = \frac{p}{m}, & \dot{\phi} & = \frac{q}{C}- \frac{\kappa}{2} \phi + \sqrt{\kappa} \phi_{\rm in}~ \\
\dot{p} & = - m \omega_0^2 \xi - \frac{G_0}{\phi_0^2 \xi_0} \phi^2 - \gamma p + F_{\rm in}, & \dot{q} & = - C \omega_a^2 \phi  - \frac{2 G_0}{\phi_0^2 \xi_0} \phi \xi - \frac{\kappa}{2} q + \sqrt{\kappa} q_{\rm in}~.
\end{aligned}
\end{equation}
The output fields are related to the input fields by the usual I/O relations
\begin{align}
\begin{split}
\label{eq:IO}
\phi_{\rm out} = \phi_{\rm in} - \sqrt{\kappa} \phi~, \quad \quad q_{\rm out} = q_{\rm in} - \sqrt{\kappa} q~.
\end{split}
\end{align}
Driving the microwave line at the LC frequency $\phi_{\rm in} \to \overline{\phi}_{\rm in} \cos (\omega_a t) + \phi_{\rm in}$, where the overlined term is the drive strength and the second term is the vacuum fluctuation around this drive, we can solve for the steady-state solution $\overline{\phi} = \overline{\phi}_{\rm in}/\sqrt{\kappa}$ to leading order in couplings and perturbations, assuming a sufficiently strong drive $|\overline{\phi}_{\rm in}| \gg \phi_{\rm in}$. Moving to the frame co-rotating with the drive (i.e., the LC circuit) and linearizing around the strong drive, we obtain the equations of motion given in the main text, viz. 
\begin{equation}
\begin{aligned}
\label{eq:EOM_linear}
\dot{\xi} & = \frac{p}{m}, & \dot{\phi} & = - \frac{\kappa}{2} \phi + \sqrt{\kappa} \phi_{\rm in}~ \\
\dot{p} & = - m \omega_0^2 \xi - \frac{G}{\phi_0 \xi_0} \phi - \gamma p + F_{\rm in}, & \dot{q} & = - \frac{2 G}{\phi_0 \xi_0} \xi - \frac{\kappa}{2} q + \sqrt{\kappa} q_{\rm in}~,
\end{aligned}
\end{equation}
where the drive-enhanced coupling is
\begin{equation}
G \equiv \sqrt{\overline{n}} G_0, \ \ \ \overline{n} \equiv \frac{|\overline{\phi}_{\rm in}|^2}{\phi_0^2 \kappa}
\end{equation}
in terms of the number $\overline{n}$ of microwave quanta circulating in the LC circuit. In this limit the equations of motion are linear and therefore easy to solve with linear response in the frequency domain.

The observable we are interested in is the output charge $q_{\rm out}$, since $q$ is the variable that gets the information about the mechanical system. From the I/O relation \eqref{eq:IO}, this means we need the solution for $\phi(\nu)$ in terms of the various input fields. The solution for the mechanical motion $z(\nu)$ is 
\begin{equation}
\xi(\nu) = \chi_m(\nu) \left[ -\frac{G}{\phi_0 \xi_0} \phi(\nu) + F_{\rm in}(\nu) \right]
\end{equation}
in terms of the response function for the mechanical motion
\begin{equation}
\chi_m(\nu) = \frac{1}{m \left[ (\omega_0^2 - \nu^2) - i \gamma \nu \right]}.
\end{equation}
Using this and the I/O relation, we obtain the solution for $q_{\rm out}(\nu)$:
\begin{equation}
\label{eq:q_out}
q_{\rm out}(\nu) =  e^{i \phi_c(\nu)} q_{\rm in}(\nu) + 2 \left( \frac{G}{\phi_0 \xi_0} \right)^2 \kappa \chi_c^2(\nu) \chi_m(\nu) \phi_{\rm in}(\nu) - 2 \left( \frac{G}{\phi_0 \xi_0} \right) \sqrt{\kappa} \chi_c(\nu) \chi_m(\nu) F_{\rm in}(\nu),
\end{equation}
where now we use the circuit (``cavity'') response function and phase shift
\begin{equation}
\chi_c(\nu) = \frac{1}{-i \nu + \kappa/2}~, \ \ \ e^{i \phi_c(\nu)} = 1 - \kappa \chi_c(\nu) = \frac{-i \nu - \kappa/2}{i \nu + \kappa/2}~.
\end{equation}

Equation \eqref{eq:q_out} shows how both any signals of interest and a variety of noise effects are encoded onto the measured output. The signal is part of $F_{\rm in}$; for a gravitational wave it is $F_{\rm in}^{\rm sig}(\nu) = m \nu^2 D h(\nu)$, where $D$ is the equilibrium distance between the loop and sphere. Each of the three terms encodes a different noise effect. Thermal noise acting on the sphere motion will also be part of $F_{\rm in}$ and couple in at order $G$. The term of order $G^0$ represents shot noise: these are the input fluctuations in $q_{\rm in}$ which transmit through the resonator circuit. The term of order $G^2$ represents back-action noise: the random drive on the sphere from microwave fluctuations in the circuit, which is then transduced back onto the circuit and eventually shows up in the output. Computing the strain noise from Eq. \eqref{eq:q_out} is straightforward. To estimate the strain from the charge data stream we divide by the appropriate coefficient, namely
\begin{equation}
h_E(\nu) = \frac{1}{\frac{2 G}{\phi_0 \xi_0} \sqrt{\kappa} \chi_c(\nu) \chi_m(\nu) m \nu^2 D} q_{\rm out}(\nu).
\end{equation}
The noise power spectrum referred to this observable is then obtained by the Weiner-Khinchin theorem, which amounts to squaring Eq. \eqref{eq:q_out} and taking the expectation value:
\begin{equation}
S_{hh} = \frac{1}{\left( \frac{ 2G}{\phi_0 \xi_0} \right)^2 \kappa |\chi_c|^2 |\chi_m|^2 m^2 \nu^4  D^2} S_{qq} + \left( \frac{G}{\phi_0 \xi_0} \right)^2 \frac{\kappa |\chi_c|^2}{m^2 \nu^4 D^2} S_{\phi\phi} + \frac{1}{m^2 \nu^4 D^2} S_{FF}.
\end{equation}
Finally, to work out the SQL, we pick a frequency $\omega_*$ at which we want to optimize the total quantum noise. We assume  vacuum input noise $S_{qq} = q_0^2/2 = 1/2 \phi_0^2$, $S_{\phi\phi}=\phi_0^2/2$. The condition for the optimal coupling $\partial S^{\rm quantum}(\omega_*)/\partial G^2 = 0$ has solution
\begin{equation}
G^2_* = \frac{\xi_0^2}{2 \kappa |\chi_c(\omega_*)|^2 |\chi_m(\omega_*)|}
\end{equation}
and the total noise power, at this frequency, reduces to the usual SQL form:
\begin{equation}
S^{\rm quantum}_{hh}(\omega_*) \to \frac{1}{2 |\chi_m(\omega_*)| m^2 \omega_*^4 D^2} \approx \frac{1}{2 m \omega_*^2 D^2},
\end{equation}
where the approximation is in the ``free-particle'' limit $\omega_* \gg \omega_0$.

\section{Flux-position coupling}\label{sec:coupling-sup}

The dimensionless coupling constant $\beta$ parametrises the sensitivity of the flux through the pick-up loop $\Phi_p$ to changes in the position of the loop; it is defined through
\begin{align}
    \frac{\partial \Phi_p}{\partial z} = \beta R^2 b_z.
    \label{eqn:appbeta}
\end{align}
In this appendix, we give a numerical derivation of its optimal value for a square, gradiometric, pick-up loop coupled to a superconducting sphere at the centre of a quadrupolar field.

For a pick-up loop oriented in the $(x,y)$ plane, the flux only depends on the $z$-component of the sphere-induced magnetic field, integrated over the loop's surface ${\Sigma_p}$, meaning
\begin{align}
    \frac{\partial \Phi_p}{\partial z} = \int_{\Sigma_p} dA \, \frac{\partial B_{\mathrm{ind},z}(x,y,D)}{\partial z},
    \label{eqn:dphidz}
\end{align}
where $\mathbf{B}_\mathrm{ind}(\mathbf{x})$ is given by Eqs. \eqref{eqn:inducedFieldExpansion} and \eqref{eqn:sphereInducedCoefficient}.

\begin{figure}[ht!]
\centering
\includegraphics[width=.48\textwidth]{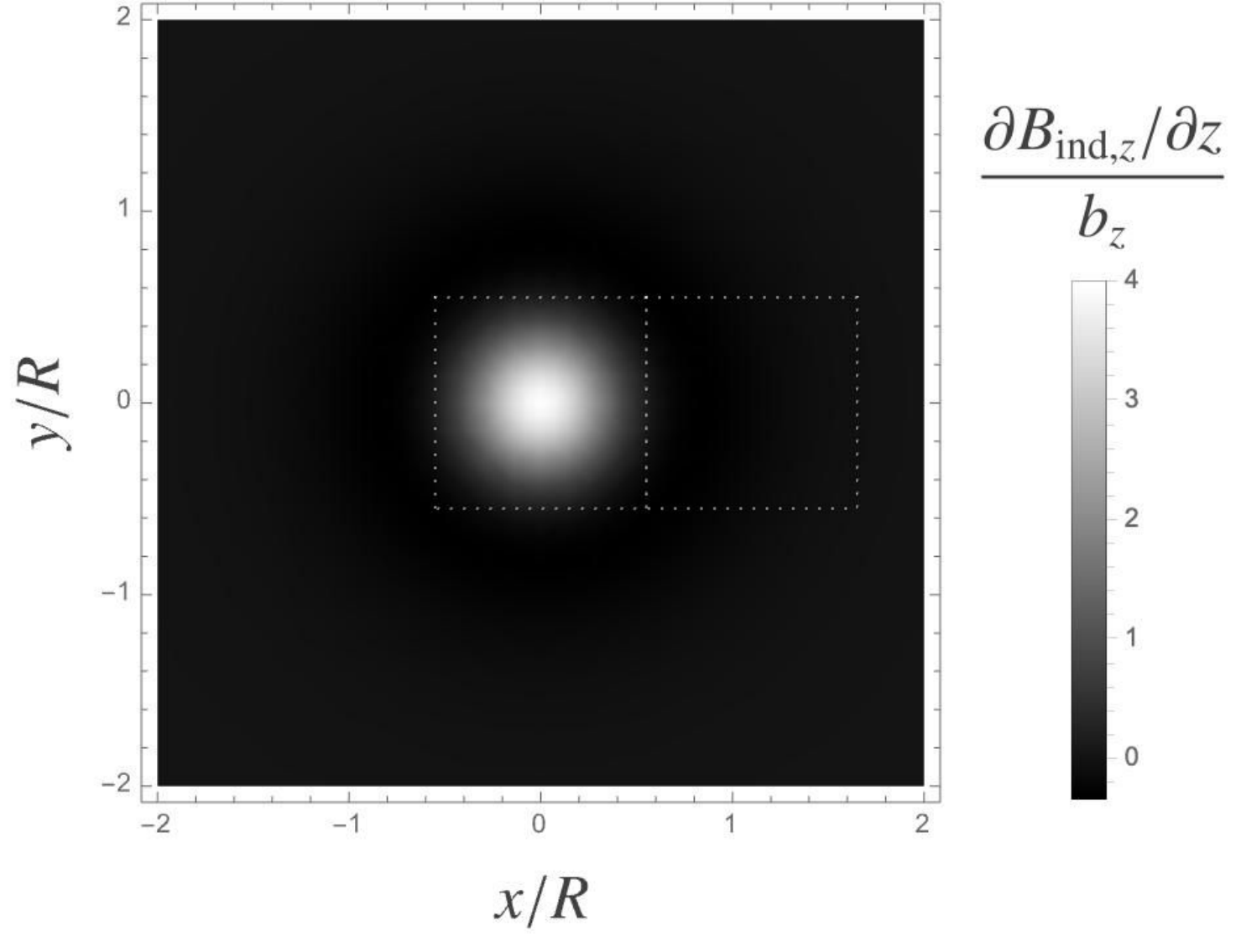}
\caption{The $z$-derivative of the  magnetic field's $z$-component, evaluated at the height of the pick-up loop, in units of $b_z$. The pick-up loop is shown in dotted white. }
\label{fig:dbdz}
\end{figure}

In Fig. \ref{fig:dbdz}, we plot $\frac{\partial B_{\mathrm{ind},z}}{\partial z}$ appearing in Eq. \eqref{eqn:dphidz}, from which we numerically evaluate the coefficient $\beta$ in Eq. \eqref{eqn:appbeta}. We find that at a pick-up loop--sphere separation $D$ equal to $R$, we achieve a maximum of $\beta \approx 1.6$ for loop of size linear size $\approx 1.1 R$. 

The change in flux due to a change in sphere--loop separation $\delta \xi = h D e^{-i\omega_g t}$ may be written as $\delta \Phi_p = h D \, \partial \Phi_p/ \partial z$, and so we find
\begin{align}
    \frac{\delta \Phi_p}{\Phi^{(0)}} \approx -2.9 h e^{-i\omega_g t}
    \label{eqn:StrainFlux}
\end{align}
for $\beta = 1.6$, where $\Phi^{(0)}$ is the unperturbed flux through the loop.

\end{document}